\documentclass[12pt]{iopart}
\usepackage{iopams}

\newcommand{\eq}[1]{(\ref{#1})}
\newcommand{\bun}{\hat{\mathbf{b}}}
\newcommand{\eun}{\hat{\mathbf{e}}}

\newcommand{\boldr}{\mathbf{r}}
\newcommand{\bv}{\mathbf{v}}

\newcommand{\bR}{\mathbf{R}}

\newcommand{\bB}{\mathbf{B}}
\newcommand{\bE}{\mathbf{E}}

\newcommand{\bW}{\mathbf{W}}

\newcommand{\matrixtop}[1]{\buildrel\leftrightarrow\over{#1}}
\newcommand{\matI}{\matrixtop{\mathbf{I}}}

\newcommand{\ssim}{ {\scriptstyle {{_{\displaystyle <}}\atop{\displaystyle \sim}}} }
\newcommand{\gsim}{ {\scriptstyle {{_{\displaystyle >}}\atop{\displaystyle \sim}}} }
\newcommand{\dotcross}{ \raise 0.65ex\hbox{${\scriptstyle {{_{\displaystyle \cdot}}\atop\times}}$} }
\newcommand{\crossdot}{ \raise 0.5ex\hbox{${\scriptstyle {{_\times}\atop{\displaystyle \cdot}}}$} }

\newcommand{\zetabf}{\mbox{\boldmath$\zeta$}}

\newcommand{\zun}{\hat{\zetabf}}

\newcommand{\sumsig}{ \raise -1.3ex\hbox{${{\displaystyle \sum}\atop{\scriptstyle \sigma}}$} }
\newcounter{appnumb}

\newcommand{\alabel}[1]{\refstepcounter{appnumb} \label{app:#1}}
\newcommand{\aref}[1]{\ref{app:#1}}

\begin{document}
\title{Sources of intrinsic rotation in the low flow ordering}
\author{Felix I Parra and Michael Barnes}
\address{Rudolf Peierls Centre for Theoretical Physics, University of Oxford, Oxford, OX1 3NP, UK}
\eads{\mailto{f.parradiaz1@physics.ox.ac.uk}}
\author{Peter J Catto}
\address{Plasma Science and Fusion Center, Massachusetts Institute of Technology, Cambridge, MA 02139, USA}

%%%%%%%%%%%%%%%%%%%%%%%%%%%%%%%%%%%%%%%%%%%%%%%%%%%%%%%%%%%%%%%%%%%%%%%%
\begin{abstract}
A low flow, $\delta f$ gyrokinetic formulation to obtain the
intrinsic rotation profiles is presented. The momentum
conservation equation in the low flow ordering contains new terms,
neglected in previous first principles formulations, that may
explain the intrinsic rotation observed in tokamaks in the absence
of external sources of momentum. The intrinsic rotation profile
depends on the density and temperature profiles and on the up-down
asymmetry.
\end{abstract}
%%%%%%%%%%%%%%%%%%%%%%%%%%%%%%%%%%%%%%%%%%%%%%%%%%%%%%%%%%%%%%%%%%%

\submitto{Nuclear Fusion}
\maketitle

%%%%%%%%%%%%%%%%%%%%%%%%%%%%%%%%%%%%%%%%%%%%%%%%%%%%%%%%%%%%%%%%
\section{Introduction} \label{sect_intro}
%%%%%%%%%%%%%%%%%%%%%%%%%%%%%%%%%%%%%%%%%%%%%%%%%%%%%%%%%%%%%%%%

Experimental observations have shown that tokamak plasmas rotate
spontaneously without momentum input \cite{rice07}. This intrinsic
rotation has been the object of recent work \cite{rice07, nave10}
because of its relevance for ITER \cite{ikeda07}, where the
projected momentum input from neutral beams is small, and the
rotation is expected to be mostly intrinsic.

The origin of the intrinsic rotation is still unclear. There has
been some theoretical work in turbulent transport of momentum
using gyrokinetic simulations \cite{peeters07, waltz07, roach09,
camenen09, camenen09b, casson09, casson10, barnes11, highcock10},
and two main mechanisms have been proposed as candidates to
explain intrinsic rotation. On the one hand, the momentum pinch
due to the Coriolis drift \cite{peeters07} has been argued to
transport momentum generated in the edge. On the other hand, it
has been argued that up-down asymmetry generates intrinsic
rotation \cite{camenen09, camenen09b}. However, neither of these
explanations are able to account for all experimental
observations. The up-down asymmetry is only large in the edge,
generating rotation in that region that then needs to be
transported inwards by the Coriolis pinch. Thus, intrinsic
rotation in the core could only be explained by the pinch. The
pinch of momentum is not sufficient because it does not allow the
toroidal rotation to change sign in the core as is observed
experimentally \cite{navepc}.

In this article we present a new model implementable in $\delta f$
flux tube simulations \cite{dorland00, candy03a, dannert05,
peeters09c}. This model is based on the low flow ordering of
\cite{parra10a}, and self-consistently includes higher order
contributions. As a result, new drive terms for the intrinsic
rotation appear that depend on the gradients of the background
profiles of density and temperature.

We recast the results from \cite{parra10a} in a form similar to
the equations in the high flow ordering \cite{artun94, sugama97}.
These are the equations that have been implemented in most
gyrokinetic codes that are employed to study momentum transport.
For this reason, the new form of the equations is useful to
identify the differences with previous models. In addition, we
discuss how the new contributions drive intrinsic rotation and we
show that the intrinsic rotation resulting from these new
processes depends on density and temperature gradients.

In the remainder of this article we present the model, developed
originally in \cite{parra10a}, in a form more suitable for $\delta
f$ flux tube simulation. In Section~\ref{sect_eq} we give the
complete model, and in Section~\ref{sect_discussion} we discuss
its implications for intrinsic rotation. Appendix~\aref{app_fi_eq}
contains the details of the transformation from the equations in
\cite{parra10a} to the formulation in this article. In
Appendix~\aref{app_pi} we discuss different forms of deriving the
radial flux of toroidal angular momentum, showing that the final
form presented here is convenient and has many advantages.

%%%%%%%%%%%%%%%%%%%%%%%%%%%%%%%%%%%%%%%%%%%%%%%%%%%%%%%%%%%%%%%%
\section{Transport of toroidal angular momentum} \label{sect_eq}
%%%%%%%%%%%%%%%%%%%%%%%%%%%%%%%%%%%%%%%%%%%%%%%%%%%%%%%%%%%%%%%%

The derivation of the transport of toroidal angular momentum in
the low flow regime, including both turbulence and neoclassical
effects, is described in detail in \cite{parra10a}. To simplify
the derivation, the extra expansion parameter $B_p/B =
\varepsilon/q \ll 1$ was employed, assuming that the turbulence
length scales and amplitudes do not depend strongly on $B_p/B$.
Here $B$ is the total magnetic field and $B_p$ its poloidal
component, $\varepsilon = r/R$ is the inverse aspect ratio of the
flux surface, $q$ is the safety factor, $r$ is the minor radius of
the flux surface and $R$ is the major radius. The ratio $B_p/B$ is
below or around 0.1 across the core in most tokamaks
($\varepsilon$ is small near the magnetic axis and $q$ is large
near the edge). In this section, we review the results of
reference~\cite{parra10a} and we recast them in a more convenient
form.

We assume that the turbulence is electrostatic and that the
magnetic field is axisymmetric, i.e., $\bB = I \nabla \zeta +
\nabla \zeta \times \nabla \psi$, where $\psi$ is the poloidal
magnetic flux, $\zeta$ is the toroidal angle, and we use a
poloidal angle $\theta$ as our third spatial coordinate. With an
axisymmetric magnetic field, in steady state and in the absence of
momentum input, the equation that determines the rotation profile
is $\langle \langle R \zun \cdot \matrixtop{\mathbf{P}}_i \cdot
\nabla \psi \rangle_\psi \rangle_\mathrm{T} = 0$, where
$\matrixtop{\mathbf{P}}_i = \int d^3v^\prime\, f_i M \bv^\prime
\bv^\prime$ is the ion stress tensor, $M$ is the ion mass, $\zun$
is the unit vector in the toroidal direction, $\langle \ldots
\rangle_\psi = (V^\prime)^{-1} \int d\theta\, d\zeta\, (...)/(\bB
\cdot \nabla \theta)$ is the flux surface average, $V^\prime
\equiv dV/d\psi = \int d\theta\, d\zeta\, (\bB \cdot \nabla
\theta)^{-1}$ is the derivative of the volume with respect to
$\psi$, and $\langle \ldots \rangle_\mathrm{T}$ is the coarse
grain or ``transport" average over the time and length scales of
the turbulence, assumed much shorter than the transport time scale
$\delta_i^{-2} a/v_{ti}$ and the minor radius $a$. Here $\delta_i
= \rho_i/a \ll 1$ is the ion gyroradius $\rho_i$ over the minor
radius of the tokamak $a$, and $v_{ti}$ is the ion thermal speed.
Note that we use the prime in $\bv^\prime$ to indicate that the
velocity is measured in the laboratory frame. Later we will find
the equations in a convenient rotating frame where the velocity is
$\bv = \bv^\prime - R\Omega_\zeta \zun$.

In reference~\cite{parra10a} we derived a method to calculate
$\langle \langle R\zun \cdot \matrixtop{\mathbf{P}}_i \cdot \nabla
\psi \rangle_\psi \rangle_\mathrm{T}$ to order $(B/B_p) \delta_i^3
p_i R |\nabla \psi|$, with $p_i$ the ion pressure. We present the
method again in a different form to make it easier to compare with
previous works in the high flow regime \cite{artun94, sugama97}.
In subsection~\ref{sub_f} we explain how we split the distribution
function and the electrostatic potential into different pieces,
and we present the equations to self-consistently obtain them. In
subsection~\ref{sub_stress} we evaluate $\langle \langle R \zun
\cdot \matrixtop{\mathbf{P}}_i \cdot \nabla \psi \rangle_\psi
\rangle_\mathrm{T}$ employing the pieces of the distribution
function and the potential obtained in subsection~\ref{sub_f}.
Before presenting all the results, we emphasize that our results
and order of magnitude estimates are valid for $\delta_i \ll B_p/B
\ll 1$, assuming that the turbulence does not scale strongly with
$B_p/B$, and for collisionality in the range $\delta_i^2 \ll
qR\nu_{ii}/v_{ti} \ssim 1$ \cite{parra10a}, where $\nu_{ii}$ is
the ion-ion collision frequency.

%%%%%%%%%%%%%%%%%%%%%%%%%%%%%%%%%%%%%%%%%%%%%%%%%%%%%%%%%%%%%%%%
\subsection{Distribution function and electrostatic potential} \label{sub_f}
%%%%%%%%%%%%%%%%%%%%%%%%%%%%%%%%%%%%%%%%%%%%%%%%%%%%%%%%%%%%%%%%

The electrostatic potential is composed to the order of interest
by the pieces in Table~\ref{table_phi} \cite{parra10a}. The
axisymmetric long wavelength pieces $\phi_0 (\psi, t)$,
$\phi_1^\mathrm{nc} (\psi, \theta, t)$ and $\phi_2^\mathrm{nc}
(\psi, \theta, t)$ are the zeroth, first and second order
equilibrium pieces of the potential. The lowest order component
$\phi_0$ is a flux surface function. The corrections
$\phi_1^\mathrm{nc}$ and $\phi_2^\mathrm{nc}$ give the electric
field parallel to the flux surface, established to force
quasineutrality at long wavelengths (the superscript
$^\mathrm{nc}$ refers to neoclassical because these are long
wavelength contributions; however, turbulence can affect the final
value of $\phi^\mathrm{nc}_2$). We need not calculate
$\phi_2^\mathrm{nc}$ because it will not appear in the final
expression for $\langle \langle R\zun \cdot
\matrixtop{\mathbf{P}}_i \cdot \nabla \psi \rangle_\psi
\rangle_\mathrm{T}$. The piece $\phi^\mathrm{tb} (\boldr, t)$ is
turbulent and includes both axisymmetric components (zonal flow)
and non-axisymmetric fluctuations. It is small in $\delta_i$ but
it has strong perpendicular gradients, i.e., $k_\bot \rho_i \sim
1$. Its parallel gradient is small, i.e., $k_{||} R \sim 1$. The
function $\phi^\mathrm{tb}$ is calculated to order $(B/B_p)
\delta_i^2 T_e/e$, i.e., $\phi^\mathrm{tb} = \phi_1^\mathrm{tb} +
\phi_2^\mathrm{tb}$ with $\phi_1^\mathrm{tb} \sim \delta_i T_e/e$
and $\phi_2^\mathrm{tb} \sim (B/B_p) \delta_i^2 T_e/e$. It is
convenient to keep both pieces together as $\phi^\mathrm{tb}$ as
we do hereafter.

\begin{table}
\caption{Pieces of the potential: $\phi = \phi_0 +
\phi_1^\mathrm{nc} + \phi_2^\mathrm{nc} + \phi^\mathrm{tb}$.}
\label{table_phi}
\begin{indented}
\item[]
\begin{tabular}{ l l l l }
\br Potential & Size & Length scales & Time scales \\
\mr $\phi_0 (\psi, t)$ & $T_e/e$ & $k a \sim 1$ & $\partial/\partial t \sim \delta_i^2 v_{ti}/a$ \\
$\phi_1^\mathrm{nc} (\psi, \theta, t)$ & $(B/B_p) \delta_i T_e/e$ & $k a \sim 1$ & $\partial/\partial t \sim \delta_i^2 v_{ti}/a$ \\
$\phi_2^\mathrm{nc} (\psi, \theta, t)$ & $(B/B_p)^2 \delta_i^2 T_e/e$ & $k a \sim 1$ & $\partial/\partial t \sim \delta_i^2 v_{ti}/a$ \\
$\phi^\mathrm{tb} (\boldr, t)$ & $\phi_1^\mathrm{tb} \sim \delta_i T_e/e$ & $k_\bot \rho_i \sim 1$ & $\partial/\partial t \sim v_{ti}/a$ \\
 & $\phi_2^\mathrm{tb} \sim (B/B_p) \delta_i^2 T_e/e$ & $k_{||} R \sim 1$  & \\
\br
\end{tabular}
\end{indented}
\end{table}

To write the distribution function it will be useful to consider
the reference frame that rotates with toroidal angular velocity
$\Omega_\zeta = - c\, \partial_\psi \phi_0$. In this new reference
frame it is easier to compare with previous formulations
\cite{artun94, sugama97}. To shorten the presentation, we perform
the change of reference frame directly in the gyrokinetic
variables. It is possible to do so easily because we are expanding
in the parameter $B/B_p \gg 1$. We first present the gyrokinetic
variables that we obtained for the laboratory frame and we argue
later how they must be modified to give the gyrokinetic variables
in the rotating frame. In \cite{parra10a} we used as gyrokinetic
variables the gyrocenter position $\bR = \boldr + \bR_1 + \bR_2 +
\ldots$, the gyrokinetic kinetic energy $E = E_0 + E_1 + E_2 +
\ldots$, the magnetic moment $\mu = \mu_0 + \mu_1 + \ldots$ and
the gyrokinetic gyrophase $\varphi = \varphi_0 + \varphi_1 +
\ldots$, where $E_0 = (v^\prime)^2/2$ is the particle kinetic
energy in the laboratory frame, $\mu_0 = (v_\bot^\prime)^2/2B$ is
the lowest order magnetic moment, $\varphi_0 = \arctan (
\bv^\prime \cdot \eun_2/\bv^\prime \cdot \eun_1)$ is the lowest
order gyrophase, $\bR_1 = \Omega_i^{-1} \bv^\prime \times \bun
\sim \delta_i a$ is the first order correction to the gyrocenter
position, $E_1 = Ze (\phi - \langle \phi \rangle_i)/M \sim
\delta_i v_{ti}^2$ is the first order correction to the
gyrokinetic kinetic energy, and the corrections $\bR_2 \sim
\delta_i^2 a$, $E_2 \sim \delta_i^2 v_{ti}^2$, $\mu_1 \sim
\delta_i v_{ti}^2/B$ and $\varphi_1 \sim \delta_i$ are defined in
\cite{parra08}. Here $\Omega_i = ZeB/Mc$ is the ion gyrofrequency,
$\eun_1(\boldr)$ and $\eun_2(\boldr)$ are two orthonormal vectors
such that $\eun_1 \times \eun_2 = \bun$, and $\langle \ldots
\rangle_i = (2\pi)^{-1} \oint d\varphi\, (\ldots)|_{\bR, E, \mu,
t}$ is the gyroaverage holding $\bR$, $E$, $\mu$ and $t$ fixed.
When the ion distribution function is written as a function of
these gyrokinetic variables, it does not depend on the gyrophase
$\varphi$ up to order $(B_p/B) \delta_i^2 (qR\nu_{ii}/v_{ti})
f_{Mi}$ \cite{parra10a, parra08}, where $f_{Mi}$ is the lowest
order distribution function that is a Maxwellian. For the magnetic
moment and the gyrophase, only the first order corrections $\mu_1$
and $\varphi_1$ are needed because the lowest order distribution
function $f_{Mi}$ does not depend on $\mu$ or $\varphi$. Moreover,
in \cite{parra10a} we expand for $1 \gg B_p/B \gg \delta_i$, and
the distribution function need only be known to order $(B/B_p)
\delta_i^2 f_{Mi}$. Consequently, the piece of the distribution
function that depends on the gyrophase, of order $(B_p/B)
\delta_i^2 (qR\nu_{ii}/v_{ti}) f_{Mi}$, is negligible, and the
gyrokinetic variables $\bR$ and $E$ only need to be obtained to
order $(B/B_p) \delta_i^2 a$ and $(B/B_p) \delta_i^2 v_{ti}^2$,
respectively, implying that the corrections $\bR_2$ and $E_2$ are
not needed for the final result. To change to the new reference
frame, where the velocity is $\bv = \bv^\prime - R \Omega_\zeta
\zun$, the distribution function that is independent of $\varphi$
has to be written as a function of the new gyrokinetic variables
$\bR$, $\varepsilon$ and $\mu$, where $\varepsilon$ is a new
variable that will be defined shortly. Note that the gyrocenter
position and the magnetic moment are the same in both reference
frames to the order of interest. In the case of $\mu$, the reason
is that $\mu$ is obtained such that its time derivative vanishes,
$d\mu/dt = 0$, making its definition unique. For $\bR$, the reason
is that the toroidal rotation has two components, one parallel to
the magnetic field, $R \Omega_\zeta \zun \cdot \bun = I
\Omega_\zeta / B \sim (B/B_p) \delta_i v_{ti}$, and the other
perpendicular, $R \Omega_\zeta |\zun - \bun \bun \cdot \zun| =
|\nabla \psi| \Omega_\zeta/ B \sim \delta_i v_{ti}$, and the
parallel velocity is larger by $B/B_p \gg 1$. Since the
gyrokinetic variable $\bR$ is to be obtained to order $(B/B_p)
\delta_i^2 a$, and in $\bR$ only the perpendicular velocity
$\bv_\bot^\prime = \bv_\bot + R\Omega_\zeta ( \zun - \bun \bun
\cdot \zun)$ enters to order $\delta_i a$, we can safely neglect
the corrections due to the change of reference frame because they
are of order $\delta_i^2 a$. In contrast, the kinetic energy $E$
as defined in \cite{parra10a} cannot be used in the rotating frame
because it includes the parallel velocity $v^\prime_{||} = v_{||}
+ I \Omega_\zeta/B$. We use a new kinetic energy variable
$\varepsilon$ that is related to the old kinetic energy variable
by $\varepsilon = E - I \Omega_\zeta u^\prime/B$, where $u^\prime
= \pm \sqrt{2(E - \mu B)}$ is the gyrokinetic parallel velocity in
the laboratory frame. It is easy to check that $u = \pm
\sqrt{2[\varepsilon - \mu B + (I/B)^2 \Omega_\zeta^2/2]}$ is equal
to $u = u^\prime - I \Omega_\zeta/B$ and it is the gyrokinetic
parallel velocity in the rotating frame. With this relation, we
find that another way to interpret the new energy variable
\begin{equation} \label{varepsilon2}
\varepsilon = \frac{u^2}{2} + \mu B - \frac{R^2 \Omega_\zeta^2}{2}
\end{equation}
is realizing that it is the kinetic energy in the rotating frame
plus the potential due to the centrifugal force. To write
expression \eq{varepsilon2} we have used that $I/B = R +
O[(B_p/B)^2 R]$ for $B_p/B \ll 1$. In Appendix~\aref{app_fi_eq} we
rewrite the results in \cite{parra10a} using the new gyrokinetic
kinetic energy $\varepsilon$.

The different pieces of the ion distribution function are given in
Table~\ref{table_fi} \cite{parra10a}. The functions $f_{Mi}$,
$H_{i1}^\mathrm{nc}$, $H_{i2}^\mathrm{nc}$ and
$H_{i2}^\mathrm{tb}$ are axisymmetric long wavelength
contributions. The Maxwellian
\begin{equation}
f_{Mi} (\psi(\bR), \varepsilon) = n_i (\psi(\bR)) \left [
\frac{M}{2\pi T_i(\psi(\bR))} \right ]^{3/2} \exp \left ( -
\frac{M \varepsilon}{T_i(\psi(\bR))} \right )
\end{equation}
is uniform on a flux surface. The first and second order
corrections $H_{i1}^\mathrm{nc}$ and $H_{i2}^\mathrm{nc}$ are
neoclassical corrections, and they are not the functions
$F_{i1}^\mathrm{nc}$ and $F_{i2}^\mathrm{nc}$ in \cite{parra10a}
because we are now working in the rotating frame. The function
$H_{i2}^\mathrm{tb}$ is an axisymmetric piece of the distribution
function that originates from collisions acting on the ions
transported by turbulent fluctuations into the volume between two
adjacent flux surfaces \cite{parra10a}. The function
$f_i^\mathrm{tb}$ is the turbulent contribution. It will be
determined self-consistently up to order $(B/B_p) \delta_i^2
f_{Mi}$, i.e., $f_i^\mathrm{tb} = f_{i1}^\mathrm{tb} +
f_{i2}^\mathrm{tb}$ with $f_{i1}^\mathrm{tb} \sim \delta_i f_{Mi}$
and $f_{i2}^\mathrm{tb} \sim (B/B_p) \delta_i^2 f_{Mi}$. It is
convenient to combine both pieces of the turbulent distribution
function into one function $f_i^\mathrm{tb}$.

\begin{table}
\caption{Pieces of the ion distribution function: $f_i = f_{Mi} +
H_{i1}^\mathrm{nc} + H_{i2}^\mathrm{nc} + H_{i2}^\mathrm{tb} +
f_i^\mathrm{tb}$.} \label{table_fi}

\begin{tabular}{ l l l l }
\br Distribution function & Size & Length scales & Time scales \\
\mr $f_{Mi} (\psi(\bR), \varepsilon, t)$ & $f_{Mi}$ & $ka \sim 1$ & $\partial/\partial t \sim \delta_i^2 v_{ti}/a$ \\
$H_{i1}^\mathrm{nc} (\psi(\bR), \theta(\bR), \varepsilon, \mu, t)$ & $(B/B_p) \delta_i f_{Mi}$ & $ka \sim 1$ & $\partial/\partial t \sim \delta_i^2 v_{ti}/a$ \\
$H_{i2}^\mathrm{nc} (\psi(\bR), \theta(\bR), \varepsilon, \mu, t)$ & $(B/B_p)^2 \delta_i^2 f_{Mi}$ & $ka \sim 1$ & $\partial/\partial t \sim \delta_i^2 v_{ti}/a$ \\
$H_{i2}^\mathrm{tb} (\psi(\bR), \theta(\bR), \varepsilon, \mu, t)$ & $(B/B_p) (v_{ti}/qR \nu_{ii}) \delta_i^2 f_{Mi}$ & $ka \sim 1$ & $\partial/\partial t \sim \delta_i^2 v_{ti}/a$ \\
$f_i^\mathrm{tb} (\bR, \varepsilon, \mu, t)$ & $f_{i1}^\mathrm{tb} \sim \delta_i f_{Mi}$ & $k_\bot \rho_i \sim 1$ & $\partial/\partial t \sim v_{ti}/a$ \\
 & $f_{i2}^\mathrm{tb} \sim (B/B_p) \delta_i^2 f_{Mi}$ & $k_{||} R \sim 1$ & \\
 \br
\end{tabular}
\end{table}

The electron distribution function is very similar to the ion
distribution function. It will have its own gyrokinetic variables
that can be easily deduced from the ion counterparts. To the order
of interest in this calculation, the electron distribution
function is determined by the pieces in Table~\ref{table_fe}. The
long wavelength, axisymmetric pieces $f_{Me}$ and
$H_{e1}^\mathrm{nc}$ are the lowest order Maxwellian and the first
order neoclassical correction. The second order long wavelength
neoclassical correction is not needed for transport of momentum
because of the small electron mass. The piece $f_e^\mathrm{tb}$ is
the short wavelength, turbulent component that will be
self-consistently calculated to order $(B/B_p) \delta_i^2 f_{Me}$.

\begin{table}
\caption{Pieces of the electron distribution function: $f_e =
f_{Me} + H_{e1}^\mathrm{nc} + f_e^\mathrm{tb}$.} \label{table_fe}

\begin{indented}
\item[]
\begin{tabular}{ l l l l }
\br Distribution function & Size & Length scales & Time scales \\
\mr $f_{Me} (\psi(\bR), \varepsilon, t)$ & $f_{Me}$ & $k a \sim 1$ & $\partial/\partial t \sim \delta_i^2 v_{ti}/a$ \\
$H_{e1}^\mathrm{nc} (\psi(\bR), \theta(\bR), \varepsilon, \mu, t)$ & $(B/B_p) \delta_i f_{Me}$ & $k a \sim 1$ & $\partial/\partial t \sim \delta_i^2 v_{ti}/a$ \\
$f_e^\mathrm{tb} (\bR, \varepsilon, \mu, t)$ & $f_{e1}^\mathrm{tb} \sim \delta_i f_{Me}$ & $k_\bot \rho_i \sim 1$ & $\partial/\partial t \sim v_{ti}/a$ \\
 & $f_{e2}^\mathrm{tb} \sim (B/B_p) \delta_i^2 f_{Me}$ & $k_{||} R \sim 1$ & \\
 \br
\end{tabular}
\end{indented}
\end{table}

We now proceed to describe how to find the different pieces of the
distribution function and the potential. We use the equations in
\cite{parra10a} but we change to the new gyrokinetic kinetic
energy $\varepsilon$. The details of this transformation are
contained in Appendix~\aref{app_fi_eq}.

%%%%%%%%%%%%%%%%%%%%%%%%%%%%%%%%%%%%%%%%%%%%%%%%%%%%%%%%%%%%%%%%%%%%%%%%%
\subsubsection{First order neoclassical distribution function and potential.}
%%%%%%%%%%%%%%%%%%%%%%%%%%%%%%%%%%%%%%%%%%%%%%%%%%%%%%%%%%%%%%%%%%%%%%%%%
The equation for $H_{i1}^\mathrm{nc}$ is
\begin{equation}
\fl u \bun \cdot \nabla_\bR \left \{ H_{i1}^\mathrm{nc} +
\frac{Ze\phi_1^\mathrm{nc}}{T_i} f_{Mi} + \left [ \frac{1}{p_i}
\frac{\partial p_i}{\partial \psi} + \left
(\frac{M\varepsilon}{T_i} - \frac{5}{2} \right ) \frac{1}{T_i}
\frac{\partial T_i}{\partial \psi} \right ] \frac{Iu
f_{Mi}}{\Omega_i} \right \} - C_{ii}^{(\ell)} \{
H_{i1}^\mathrm{nc} \} = 0, \label{eqHi1nc}
\end{equation}
where $u = \pm \sqrt{2(\varepsilon - \mu B + R^2
\Omega_\zeta^2/2)} \simeq \pm \sqrt{2(\varepsilon - \mu B)}$ is
the gyrokinetic parallel velocity and $C_{ii}^{(\ell)}$ is the
linearized ion-ion collision operator. The correction
$H_{i1}^\mathrm{nc}$ gives the parallel component of the velocity
\cite{hinton76, helander02bk} $n_i \bW_i^\mathrm{nc} = \bun \int
d^3v\, H_{i1}^\mathrm{nc} v_{||} = - (c I \bun /Ze B)\partial_\psi
p_i + (k n_i cI \bB/Ze \langle B^2 \rangle_\psi)\partial_\psi
T_i$, where $k$ is a flux function that depends on the
collisionality and the magnetic geometry.

The equation for $H_{e1}^\mathrm{nc}$ is similar to \eq{eqHi1nc}
and it is given by \cite{hinton76, helander02bk}
\begin{eqnarray} \label{eqHe1nc}
\fl u \bun \cdot \nabla_\bR \left \{ H_{e1}^\mathrm{nc} -
\frac{e\phi_1^\mathrm{nc}}{T_e} f_{Me} - \left [\frac{1}{p_e}
\frac{\partial p_e}{\partial \psi} + \left
(\frac{M\varepsilon}{T_e} - \frac{5}{2} \right ) \frac{1}{T_e}
\frac{\partial T_e}{\partial \psi} \right ] \frac{Iu
f_{Me}}{\Omega_e} \right \} \nonumber\\ - C_{ee}^{(\ell)} \{
H_{e1}^\mathrm{nc} \} - C_{ei}^{(\ell)} \{ H_{e1}^\mathrm{nc} \} =
- \frac{e f_{Me}}{T_e} u \bun \cdot \bE^A,
\end{eqnarray}
where $m$ and $\Omega_e = eB/mc$ are the electron mass and
gyrofrequency, $\bE^A$ is the electric field driven by the
transformer, $C_{ee}^{(\ell)}$ is the linearized electron-electron
collision operator and $C_{ei}^{(\ell)}$ is the linearized
electron-ion collision operator. The lowest order solution for
$H_{e1}^\mathrm{nc}$ is the Maxwell-Boltzmann response
$(e\phi_1^\mathrm{nc}/T_e) f_{Me} \sim (B/B_p)\delta_i f_{Me}$.
The rest of the terms are small because they are of order $(B/B_p)
\delta_e f_{Me} \sim (B/B_p) \sqrt{m/M} \delta_i f_{Mi} \ll
(B/B_p) \delta_i f_{Me}$, where $\delta_e = \rho_e/a$ is the ratio
between the electron gyroradius $\rho_e$ and the minor radius $a$.

Finally the poloidal variation of the potential is determined by
quasineutrality,
\begin{equation}
Z\int d^3v\, H_{i1}^\mathrm{nc} = \frac{e\phi_1^\mathrm{nc}}{T_e}
n_e, \label{eqphi1nc}
\end{equation}
giving $e\phi_1^\mathrm{nc}/T_e \sim (B/B_p) \delta_i$.

%%%%%%%%%%%%%%%%%%%%%%%%%%%%%%%%%%%%%%%%%%%%%%%%%%%%%%%%%%%%%%%%%%%%%%%%%
\subsubsection{Turbulent distribution function and potential.}
%%%%%%%%%%%%%%%%%%%%%%%%%%%%%%%%%%%%%%%%%%%%%%%%%%%%%%%%%%%%%%%%%%%%%%%%%

The turbulent piece of the ion distribution function is obtained
using the gyrokinetic equation (see Appendix \aref{app_fi_eq})
\begin{eqnarray}
\fl \frac{D f_i^\mathrm{tb}}{D t} + & \left ( u \bun + \bv_M +
\bv_C + \bv_{E1}^\mathrm{nc} + \bv_E^\mathrm{tb} \right ) \cdot
\nabla_\bR f_i^\mathrm{tb} - \left \langle C_{ii}^{(\ell)} \left
\{ h_i^\mathrm{tb} \right \} \right \rangle_i \nonumber\\ \fl & =
- \bv_E^\mathrm{tb} \cdot \nabla_\bR \psi \left [\frac{1}{n_i}
\frac{\partial n_i}{\partial \psi} + \left (
\frac{M\varepsilon}{T_i} - \frac{3}{2} \right ) \frac{1}{T_i}
\frac{\partial T_i}{\partial \psi} + \frac{MIu}{BT_i}
\frac{\partial \Omega_\zeta}{\partial \psi} \right ] f_{Mi} -
\bv_E^\mathrm{tb} \cdot \nabla_\bR H_{i1}^\mathrm{nc} \nonumber
\\\fl & - \frac{Ze f_{Mi}}{T_i} \left ( u \bun + \bv_M + \bv_C
\right ) \cdot \nabla_\bR \langle \phi^\mathrm{tb} \rangle_i +
\frac{Ze}{M} \frac{\partial H_{i1}^\mathrm{nc}}{\partial
\varepsilon} \left ( u \bun + \bv_M \right ) \cdot \nabla_\bR
\langle \phi^\mathrm{tb} \rangle_i, \label{eqfitb}
\end{eqnarray}
where $D/Dt = \partial_t + R \Omega_\zeta \zun \cdot \nabla_\bR$
is the time derivative in the rotating frame, $u = \pm
\sqrt{2[\varepsilon - \mu B + R^2 \Omega_\zeta^2/2]} \simeq \pm
\sqrt{2(\varepsilon - \mu B)}$ is the parallel velocity in the
rotating frame, $\bv_M = (\mu/\Omega_i) \bun \times \nabla_\bR B +
(u^2/\Omega_i) \bun \times (\bun \cdot \nabla_\bR \bun)$ are the
$\nabla B$ and curvature drifts, $\bv_C = (2 u
\Omega_\zeta/\Omega_i) \bun \times [(\nabla R \times \zun) \times
\bun]$ is the Coriolis drift, $\bv_{E1}^\mathrm{nc} = - (c/B)
\nabla_\bR \phi_1^\mathrm{nc} \times \bun$ and $\bv_E^\mathrm{tb}
= - (c/B) \nabla_\bR \langle \phi^\mathrm{tb} \rangle_i \times
\bun$ are the neoclassical and turbulent $\bE \times \bB$ drifts,
$C_{ii}^{(\ell)} \{ h_i^\mathrm{tb} \}$ is the linearized ion-ion
collision operator, and $\langle \ldots \rangle_i = (2\pi)^{-1}
\oint d\varphi\, (\ldots)|_{\bR, E, \mu, t}$ is the gyroaverage
holding the ion gyrokinetic variables $\bR = \boldr +
\Omega_i^{-1} \bv \times \bun + \ldots$, $E$, $\mu$ and $t$ fixed.
The function that enters in the collision operator is
\begin{equation}
h_i^\mathrm{tb} = f_{ig}^\mathrm{tb} + \frac{Ze (\phi^\mathrm{tb}
- \langle \phi^\mathrm{tb} \rangle_i)}{M} \left ( -
\frac{Mf_{Mi,0}}{T_i} + \frac{\partial
H_{i1,0}^\mathrm{nc}}{\partial \varepsilon_0} + \frac{1}{B}
\frac{\partial H_{i1,0}^\mathrm{nc}}{\partial \mu_0} \right ).
\end{equation}
Here the subscript $_g$ in $f_{ig}^\mathrm{tb} = f_i^\mathrm{tb}
(\bR_g, v^2/2, v_\bot^2/2B, t)$ indicates that we have replaced
the variables $\bR$, $\varepsilon$ and $\mu$ by $\bR_g = \boldr +
\Omega_i^{-1} \bv \times \bun$, $v^2/2$ and $v_\bot^2/2B$;
similarly, the subscript $_0$ in $f_{Mi,0} = f_{Mi} (\psi(\boldr),
v^2/2, t)$ and $H_{i1,0}^\mathrm{nc} = H_{i1}^\mathrm{nc}
(\psi(\boldr), \theta(\boldr), v^2/2, v_\bot^2/2B, t)$ indicates
that we have replaced the variables $\bR$, $\varepsilon$ and $\mu$
by $\boldr$, $v^2/2$ and $v_\bot^2/2B$.

The equation for electrons is of the same form as the one for the
ions, giving
\begin{eqnarray}
\fl \frac{D f_e^\mathrm{tb}}{D t} + \left ( u \bun + \bv_M +
\bv_E^\mathrm{tb} \right ) \cdot \nabla_\bR f_e^\mathrm{tb} -
\left \langle C_{ee}^{(\ell)} \left \{ h_e^\mathrm{tb} \right \}
\right \rangle_e - \left \langle C_{ei}^{(\ell)} \left \{
h_e^\mathrm{tb}, h_i^\mathrm{tb} \right \} \right \rangle_e
\nonumber\\ = - \bv_E^\mathrm{tb} \cdot \nabla_\bR \psi \Bigg [
\frac{1}{n_e} \frac{\partial n_e}{\partial \psi} + \left (
\frac{M\varepsilon}{T_e} - \frac{3}{2} \right ) \frac{1}{T_e}
\frac{\partial T_e}{\partial \psi} \Bigg ] f_{Me} \nonumber\\ +
\frac{e f_{Me}}{T_e} \left ( u \bun + \bv_M \right ) \cdot
\nabla_\bR \langle \phi^\mathrm{tb} \rangle_e, \label{eqfetb}
\end{eqnarray}
where $\bv_M = - (\mu/\Omega_e) \bun \times \nabla_\bR B -
(u^2/\Omega_e) \bun \times (\bun \cdot \nabla_\bR \bun)$ are the
$\nabla B$ and curvature drifts for electrons, $\bv_E^\mathrm{tb}
= - (c/B) \nabla_\bR \langle \phi^\mathrm{tb} \rangle_e \times
\bun$ is the turbulent $\bE \times \bB$ drift, $C_{ee}^{(\ell)} \{
h_e^\mathrm{tb} \}$ is the linearized electron-electron collision
operator, $C_{ei}^\mathrm{(\ell)} \{ h_e^\mathrm{tb},
h_i^\mathrm{tb} \}$ is the linearized electron-ion collision
operator, and $\langle \ldots \rangle_e = (2\pi)^{-1} \oint
d\varphi\, (\ldots)|_{\bR, E, \mu, t}$ is the gyroaverage holding
the electron gyrokinetic variables $\bR = \boldr - \Omega_e^{-1}
\bv \times \bun + \ldots$, $E$, $\mu$ and $t$ fixed. The electron
distribution function that enters in the collision operator is
\begin{equation}
h_e^\mathrm{tb} = f_{eg}^\mathrm{tb} + \frac{e (\phi^\mathrm{tb} -
\langle \phi^\mathrm{tb} \rangle_e)}{T_e} f_{Me,0}.
\end{equation}
The subscript $_g$ on $f_{eg}^\mathrm{tb} = f_e^\mathrm{tb}
(\bR_g, v^2/2, v_\bot^2/2B, t)$ indicates that we have replaced
the variables $\bR$, $\varepsilon$ and $\mu$ by $\bR_g = \boldr -
\Omega_e^{-1} \bv \times \bun$, $v^2/2$ and $v_\bot^2/2B$;
similarly, the subscript $_0$ on $f_{Me,0} = f_{Me} (\psi(\boldr),
v^2/2, t)$ indicates that we have replaced the variables $\bR$,
$\varepsilon$ and $\mu$ by $\boldr$, $v^2/2$ and $v_\bot^2/2B$. If
we were to neglect the effect of the trapped electrons, the
solution to this equation for time and length scales typical of
ion turbulence would simply be the adiabatic response
$f_e^\mathrm{tb} \simeq (e\langle \phi^\mathrm{tb}\rangle /T_e)
f_{Me}$ because of the high parallel speed of the electrons.

Finally, the electrostatic potential $\phi^\mathrm{tb}$ is
obtained from the quasineutrality equation
\begin{eqnarray}
\fl \int d^3v\, \frac{Z^2 e (\phi^\mathrm{tb} - \langle
\phi^\mathrm{tb} \rangle_i )}{M} \left [ \frac{Mf_{Mi,0}}{T_i} -
\frac{\partial H_{i1,0}^\mathrm{nc}}{\partial \varepsilon_0} -
\frac{1}{B} \frac{\partial H_{i1,0}^\mathrm{nc}}{\partial \mu_0}
\right ] + \int d^3v\, \frac{e (\phi^\mathrm{tb} - \langle
\phi^\mathrm{tb} \rangle_e )}{T_e} f_{Me,0} \nonumber\\ = Z \int
d^3v\, f_{ig}^\mathrm{tb} - \int d^3v\, f_{eg}^\mathrm{tb}.
\label{eqphitb}
\end{eqnarray}

%%%%%%%%%%%%%%%%%%%%%%%%%%%%%%%%%%%%%%%%%%%%%%%%%%%%%%%%%%%%%%%%%%%%%%%%%
\subsubsection{Second order, long wavelength distribution function.}
%%%%%%%%%%%%%%%%%%%%%%%%%%%%%%%%%%%%%%%%%%%%%%%%%%%%%%%%%%%%%%%%%%%%%%%%%

The long wavelength pieces $H_{i2}^\mathrm{nc}$ and
$H_{i2}^\mathrm{tb}$ are given by
\begin{eqnarray}
\fl u \bun \cdot \nabla_\bR H_{i2}^\alpha - C_{ii}^{(\ell)} \{
H_{i2}^\alpha \} = \mathcal{S}^\alpha - \Bigg \langle \int d^3v\,
\mathcal{S}^\alpha \nonumber\\ + \left (
\frac{2M\varepsilon}{3T_i} - 1 \right ) \int d^3v\,
\mathcal{S}^\alpha \left ( \frac{M\varepsilon}{T_i} - \frac{3}{2}
\right ) \Bigg \rangle_\psi \frac{f_{Mi}}{n_i}, \label{eqF2}
\end{eqnarray}
where $\alpha = \mathrm{nc},\; \mathrm{tb}$. In the preceding,
\begin{eqnarray} \label{Snc}
\fl \mathcal{S}^\mathrm{nc} = & - \left [ \frac{1}{p_i}
\frac{\partial p_i}{\partial \psi} + \left (
\frac{M\varepsilon}{T_i} - \frac{5}{2} \right ) \frac{1}{T_i}
\frac{\partial T_i}{\partial \psi} \right ] f_{Mi} \left ( \bv_C -
\frac{c}{B} \nabla_\bR \phi_1^\mathrm{nc} \times \bun \right )
\cdot \nabla_\bR \psi \nonumber \\ \fl & - \frac{M I u f_{Mi}}{B
T_i} \frac{\partial \Omega_\zeta}{\partial \psi} \bv_M \cdot
\nabla_\bR \psi - \bv_M \cdot \nabla_\bR H_{i1}^\mathrm{nc} -
\frac{Ze f_{Mi}}{T_i} \left ( u \bun \cdot \nabla_\bR
\phi_2^\mathrm{nc} + \bv_M \cdot \nabla_\bR \phi_1^\mathrm{nc}
\right ) \nonumber \\ \fl & + \frac{Ze}{M} \frac{\partial
H_{i1}^\mathrm{nc}}{\partial \varepsilon} u \bun \cdot \nabla_\bR
\phi_1^\mathrm{nc} + C_{ii}^{(n\ell)} \{ H_{i1}^\mathrm{nc},
H_{i1}^\mathrm{nc} \},
\end{eqnarray}
with $C_{ii}^{(n\ell)}$ the full bilinear ion-ion collision
operator, and
\begin{eqnarray} \label{Stb}
\fl \mathcal{S}^\mathrm{tb} = - \frac{|u|}{B} \nabla_\bR \cdot
\left ( \frac{B}{|u|} \left \langle f_i^\mathrm{tb}
\bv_E^\mathrm{tb} \right \rangle_\mathrm{T} \right ) +
\frac{Ze}{M} \frac{|u|}{B} \frac{\partial}{\partial \varepsilon}
\left ( \frac{B}{|u|} \left \langle f_i^\mathrm{tb} \left ( u \bun
+ \bv_M \right ) \cdot \nabla_\bR \langle \phi^\mathrm{tb} \rangle
\right \rangle_\mathrm{T} \right ).
\end{eqnarray}

%%%%%%%%%%%%%%%%%%%%%%%%%%%%%%%%%%%%%%%%%%%%%%%%%%%%%%%%%%%%%%%%
\subsection{Calculation of the momentum transport} \label{sub_stress}
%%%%%%%%%%%%%%%%%%%%%%%%%%%%%%%%%%%%%%%%%%%%%%%%%%%%%%%%%%%%%%%%

The radial transport of toroidal angular momentum $\langle \langle
R \zun \cdot \matrixtop{\mathbf{P}}_i \cdot \nabla \psi
\rangle_\psi \rangle_\mathrm{T}$ is given in equation~(39) of
\cite{parra10a} that we reproduce here for convenience
\begin{eqnarray} \label{pi1}
\fl \langle \langle R \zun \cdot \matrixtop{\mathbf{P}}_i \cdot
\nabla \psi \rangle_\psi \rangle_\mathrm{T} = Mc \left \langle
\left \langle \frac{\partial \phi}{\partial \zeta} \int
d^3v^\prime\, f_i R (\bv^\prime \cdot \zun) \right \rangle_\psi
\right \rangle_\mathrm{T} + \frac{Mc \langle R^2
\rangle_\psi}{2Ze} \frac{\partial p_i}{\partial t} \nonumber\\+
\frac{M^2 c^2}{2 Z e} \frac{1}{V^\prime} \frac{\partial}{\partial
\psi} V^\prime \left \langle \left \langle \frac{\partial
\phi}{\partial \zeta} \int d^3v^\prime\, f_i R^2 (\bv^\prime \cdot
\zun)^2 \right \rangle_\psi \right \rangle_\mathrm{T} \nonumber\\-
\frac{M^2 c}{2Ze} \left \langle \int d^3v^\prime\, \left \langle
C_{ii} \{ f_i \} \right \rangle_\mathrm{T} R^2 (\bv^\prime \cdot
\zun)^2 \right \rangle_\psi \nonumber\\- \frac{M^3 c^2}{6Z^2e^2}
\frac{1}{V^\prime} \frac{\partial}{\partial \psi} V^\prime \left
\langle \int d^3v^\prime\, \left \langle C_{ii} \{ f_i \} \right
\rangle_\mathrm{T} R^3 (\bv^\prime \cdot \zun)^3 \right
\rangle_\psi.
\end{eqnarray}
This expression is derived in \cite{parra10a}. In
Appendix~\aref{app_pi}, we present an alternative proof that makes
clear the convenience of using form \eq{pi1}.

\begin{table}
\caption{Contributions to transport of momentum.} \label{table_Pi}

\begin{tabular}{ l l l }
\br $\Pi$ & Size [$(B/B_p)\delta_i^3 p_i R |\nabla \psi|$] & Dependences \\
\mr $\Pi_{-1}^\mathrm{tb}$ & $(B_p/B) \Delta_{ud} \delta_i^{-1}$ for $\Delta_{ud} \gsim (B/B_p) \delta_i$ & $\partial_\psi \Omega_\zeta, \Omega_\zeta, \Delta_{ud}, \partial_\psi T_i, \partial_\psi n_e, \partial_\psi T_e, \partial_\psi^2 T_i, \partial_\psi^2 n_e$ \\
 & $1$ for $\Delta_{ud} \ssim (B/B_p) \delta_i$ & \\
$\Pi_{0}^\mathrm{tb}$ & $1$ & $\partial_\psi T_i, \partial_\psi n_e, \partial_\psi T_e, \partial_\psi^2 T_i,  \partial_\psi^2 n_e, \partial_\psi^2 T_e$ \\
$\Pi_{-1}^\mathrm{nc}$ & $\Delta_{ud} (qR\nu_{ii}/v_{ti}) \delta_i^{-1}$ for $\Delta_{ud} \gsim (B/B_p) \delta_i$ & $\partial_\psi \Omega_\zeta, \Delta_{ud}, \partial_\psi T_i, \partial_\psi T_e, \partial_\psi n_e, \partial_\psi^2 T_i, \partial_\psi^2 n_e$ \\
 & $(B/B_p) (qR\nu_{ii}/v_{ti})$ for $\Delta_{ud} \ssim (B/B_p) \delta_i$ & \\
$\Pi_{0}^\mathrm{nc}$ & $(B/B_p) (qR\nu_{ii}/v_{ti})$ &
$\partial_\psi T_i, \partial_\psi n_e, \partial_\psi T_e, \partial_\psi^2 T_i, \partial_\psi^2 n_e$ \\
\br
\end{tabular}
\end{table}

Using that for $B/B_p \gg 1$, $R \bv \cdot \zun \simeq I
v_{||}/B$, and employing the decomposition of the ion distribution
function in subsection~\ref{sub_f}, we find that \eq{pi1} can be
rewritten as
\begin{equation}
\langle \langle R \zun \cdot \matrixtop{\mathbf{P}}_i \cdot \nabla
\psi \rangle_\psi \rangle_\mathrm{T} = \Pi_{-1}^\mathrm{tb} +
\Pi_0^\mathrm{tb} + \Pi_{-1}^\mathrm{nc} + \Pi_0^\mathrm{nc} +
\frac{Mc\langle R^2 \rangle_\psi}{2Ze} \frac{\partial
p_i}{\partial t}, \label{pi_final}
\end{equation}
with
\begin{equation}
\fl \Pi_{-1}^\mathrm{tb} = Mc \left \langle \left \langle
\frac{\partial \phi^\mathrm{tb}}{\partial \zeta} \int d^3v\,
f_{ig}^\mathrm{tb} \left ( \frac{I v_{||}}{B} + R \Omega_\zeta
\right ) \right \rangle_\psi \right \rangle_\mathrm{T},
\end{equation}
\begin{eqnarray}
\fl \Pi_0^\mathrm{tb} = \frac{M^2 c^2}{2Ze} \frac{1}{V^\prime}
\frac{\partial}{\partial \psi} V^\prime \left \langle \left
\langle \frac{\partial \phi^\mathrm{tb}}{\partial \zeta} \int
d^3v\, f_{ig}^\mathrm{tb} \frac{I^2 v_{||}^2}{B^2} \right
\rangle_\mathrm{T} \right \rangle_\psi \nonumber\\- \frac{M^2
c}{2Ze} \left \langle \int d^3v\, C_{ii}^{(\ell)} \{
H_{i2,0}^\mathrm{tb} \} \frac{I^2 v_{||}^2}{B^2} \right
\rangle_\psi,
\end{eqnarray}
\begin{equation}
\fl \Pi_{-1}^\mathrm{nc} = - \frac{M^2 c}{2Ze} \left \langle \int
d^3v\, C_{ii}^{(\ell)} \{ H_{i1,0}^\mathrm{nc} +
H_{i2,0}^\mathrm{nc} \} \frac{I^2 v_{||}^2}{B^2} \right
\rangle_\psi
\end{equation}
and
\begin{eqnarray}
\fl \Pi_0^\mathrm{nc} =  - \frac{M^2 c}{2Ze} \left \langle \int
d^3v\, C_{ii}^{(n\ell)} \{ H_{i1,0}^\mathrm{nc},
H_{i1,0}^\mathrm{nc} \} \frac{I^2 v_{||}^2}{B^2} \right
\rangle_\psi \nonumber\\ - \frac{M^3c^2}{6Z^2 e^2}
\frac{1}{V^\prime} \frac{\partial}{\partial \psi} V^\prime \left
\langle \int d^3v\, C_{ii}^{(\ell)} \{ H_{i1,0}^\mathrm{nc} \}
\frac{I^3 v_{||}^3}{B^3} \right \rangle_\psi.
\end{eqnarray}
Recall that the subscript $_g$ indicates that $\bR$, $\varepsilon$
and $\mu$ have been replaced by $\bR_g = \boldr + \Omega_i^{-1}
\bv \times \bun$, $v^2/2$ and $v_\bot^2/2B$, and the subscript
$_0$ that they have been replaced by $\boldr$, $v^2/2$ and
$v_\bot^2/2B$. In Table~\ref{table_Pi} we summarize the size of
all these contributions compared to the reference size $(B/B_p)
\delta_i^3 p_i R |\nabla \psi|$, and we write what they depend on.
To obtain these dependences, we use equations~\eq{eqHi1nc},
\eq{eqHe1nc}, \eq{eqphi1nc}, \eq{eqfitb}, \eq{eqfetb},
\eq{eqphitb} and \eq{eqF2}. The size estimates are taken from
\cite{parra10a}, where the turbulence was assumed to not scale
strongly with $B_p/B$. We use $\Delta_{ud}$ to denote a measure of
the flux surface up-down asymmetry. It ranges from zero for
perfect up-down symmetry to one for extreme asymmetry. Notice that
for extreme up-down asymmetry, $\Pi_{-1}^\mathrm{tb}$ and
$\Pi_{-1}^\mathrm{nc}$ clearly dominate.

%%%%%%%%%%%%%%%%%%%%%%%%%%%%%%%%%%%%%%%%%%%%%%%%%%%%%%%%%%%%%%%%%%%%%%%%%
\section{Discussion} \label{sect_discussion}
%%%%%%%%%%%%%%%%%%%%%%%%%%%%%%%%%%%%%%%%%%%%%%%%%%%%%%%%%%%%%%%%%%%%%%%%%

We finish by showing how this new formalism gives a plausible
model for intrinsic rotation. Until now, models have only
considered the contribution $\Pi_{-1}^\mathrm{tb}$, with
$f_i^\mathrm{tb}$ and $\phi^\mathrm{tb}$ obtained by employing
equations~\eq{eqfitb} and \eq{eqphitb} without the terms that
contain $H_{i1}^\mathrm{nc}$ and $\phi_1^\mathrm{nc}$. This is
acceptable for $R\Omega_\zeta \gg (B/B_p) \delta_i v_{ti}$ or high
up-down asymmetry $\Delta_{ud} \gg (B/B_p) \delta_i$. In these
limits, $\Pi_{-1}^\mathrm{tb} (\partial_\psi \Omega_\zeta,
\Omega_\zeta) \simeq - \nu^\mathrm{tb} \partial_\psi \Omega_\zeta
- \Gamma^\mathrm{tb} \Omega_\zeta + \Pi^\mathrm{tb}_{ud}$. To
obtain this last expression we have linearized around
$\partial_\psi \Omega_\zeta = 0$ and $\Omega_\zeta = 0$ for
$R\Omega_\zeta/v_{ti} \ll 1$. Here $\nu^\mathrm{tb}$ is the
turbulent diffusivity, $\Gamma^\mathrm{tb}$ is the turbulent pinch
of momentum and $\Pi^\mathrm{tb}_{ud} \sim \Delta_{ud} \delta_i^2
p_i R |\nabla \psi|$ is the value of $\Pi_{-1}^\mathrm{tb}$ at
$\Omega_\zeta = 0$ and $\partial_\psi \Omega_\zeta = 0$, and is
zero for perfect up-down asymmetry when equations~\eq{eqfitb},
\eq{eqfetb} and \eq{eqphitb} are solved without the terms that
contain $H_{i1}^\mathrm{nc}$ and $\phi_1^\mathrm{nc}$
\cite{parra11}. Notice then that imposing $\langle \langle R \zun
\cdot \matrixtop{\mathbf{P}}_i \cdot \nabla \psi \rangle_\psi
\rangle_\mathrm{T} \simeq \Pi^\mathrm{tb} = -
\nu^\mathrm{tb}\partial_\psi \Omega_\zeta - \Gamma^\mathrm{tb}
\Omega_\zeta + \Pi^\mathrm{tb}_{ud} = 0$ gives intrinsic rotation
only for up-down asymmetry or if momentum is pinched into the core
from the edge.

The complete model described in this article includes
contributions that have not been considered before. On the one
hand, the gyrokinetic equations \eq{eqfitb} and \eq{eqphitb} have
new terms depending on $H_{i1}^\mathrm{nc}$ and
$\phi_1^\mathrm{nc}$, giving $\Pi_{-1}^\mathrm{tb} \simeq -
\nu^\mathrm{tb}
\partial_\psi \Omega_\zeta - \Gamma^\mathrm{tb} \Omega_\zeta +
\Pi^\mathrm{tb}_{ud} + \Pi^\mathrm{tb}_{-1,0}$, where
$\Pi^\mathrm{tb}_{-1,0} \sim (B/B_p) \delta_i^3 p_i R |\nabla
\psi|$ is a new contribution due to the new terms in the
gyrokinetic equation. In addition, there are the new terms
$\Pi_0^\mathrm{tb}$, $\Pi^\mathrm{nc}_{-1}$ and
$\Pi_0^\mathrm{nc}$. As we did for $\Pi_{-1}^\mathrm{tb}$, we can
linearize $\Pi_{-1}^\mathrm{nc} (\partial_\psi \Omega_\zeta)$
around $\partial_\psi \Omega_\zeta = 0$ to find
$\Pi_{-1}^\mathrm{nc} \simeq - \nu^\mathrm{nc}
\partial_\psi \Omega_\zeta + \Pi^\mathrm{nc}_{ud} +
\Pi^\mathrm{nc}_{-1,0}$, where $\Pi^\mathrm{nc}_{ud} \sim
\Delta_{ud} (B/B_p) (q R \nu_{ii}/v_{ti}) \delta_i^2 p_i R |\nabla
\psi|$ and $\Pi^\mathrm{nc}_{-1,0} \sim (B/B_p)^2 (q R
\nu_{ii}/v_{ti}) \delta_i^3 p_i R |\nabla \psi|$. Combining all
these results and imposing that $\langle \langle R \zun \cdot
\matrixtop{\mathbf{P}}_i \cdot \nabla \psi \rangle_\psi
\rangle_\mathrm{T} = 0$, we obtain
\begin{eqnarray}
\fl \Omega_\zeta =  - \int_\psi^{\psi_a} d\psi^\prime\, \left.
\frac{\Pi^\mathrm{int}}{\nu^\mathrm{tb} + \nu^\mathrm{nc}} \right
|_{\psi = \psi^\prime}  \exp \left ( \int_\psi^{\psi^\prime}
d\psi^{\prime \prime}\, \left.
\frac{\Gamma^\mathrm{tb}}{\nu^\mathrm{tb} + \nu^\mathrm{nc}}
\right |_{\psi = \psi^{\prime \prime}} \right ) \nonumber \\ +
\Omega_\zeta |_{\psi = \psi_a} \exp \left ( \int_{\psi}^{\psi_a}
d\psi^\prime\, \left. \frac{\Gamma^\mathrm{tb}}{\nu^\mathrm{tb} +
\nu^\mathrm{nc}} \right |_{\psi = \psi^\prime} \right ),
\end{eqnarray}
where $\psi_a$ is the poloidal flux at the edge, $\Omega_\zeta
|_{\psi = \psi_a}$ is the rotation velocity in the edge and
$\Pi^\mathrm{int} = \Pi_{ud}^\mathrm{tb} + \Pi^\mathrm{tb}_{-1,0}
+ \Pi^\mathrm{tb}_0 + \Pi_{ud}^\mathrm{nc} +
\Pi^\mathrm{nc}_{-1,0} + \Pi^\mathrm{nc}_0$. Notice that this
equation gives a rotation profile that depends on
$\Pi^\mathrm{int}$ that in turn depends on the gradients of
temperature and density, and the magnetic field geometry. The
typical size of the rotation is $\Omega_\zeta \sim (B/B_p)
\delta_i v_{ti}/R$ for $\Delta_{ud} \ssim (B/B_p) \delta_i$ and
$\Omega_\zeta \sim \Delta_{ud} v_{ti}/R$ for $\Delta_{ud} \gsim
(B/B_p) \delta_i$.

This new model for intrinsic rotation has been constructed such
that the pinch and the up-down symmetry drive, discovered in the
high flow ordering, are naturally included. By transforming to the
frame rotating with $\Omega_\zeta$ we have made this property
explicit.

\ack Work supported in part by the post-doctoral fellowship
programme of the UK EPSRC, by the Junior Research Fellowship
programme of Christ Church at University of Oxford, by the U.S.
Department of Energy Grant No. DE-FG02-91ER-54109 at the Plasma
Science and Fusion Center of the Massachusetts Institute of
Technology, by the Center for Multiscale Plasma Dynamics of
University of Maryland and by the Leverhulme network for
Magnetised Turbulence in Astrophysical and Fusion Plasmas.

\appendix

%%%%%%%%%%%%%%%%%%%%%%%%%%%%%%%%%%%%%%%%%%%%%%%%%%%%%%%%%%%%%%%%%%%%%
\section{Equation for the distribution function in the rotating frame} \alabel{app_fi_eq}
%%%%%%%%%%%%%%%%%%%%%%%%%%%%%%%%%%%%%%%%%%%%%%%%%%%%%%%%%%%%%%%%%%%%%

In this Appendix we derive equations \eq{eqHi1nc}, \eq{eqfitb} and
\eq{eqF2} for the different pieces of the ion distribution
function, equations \eq{eqHe1nc} and \eq{eqfetb} for the different
pieces of the electron distribution function, and equations
\eq{eqphi1nc} and \eq{eqphitb} for the different pieces of the
potential. These equations are valid in the frame rotating with
angular velocity $\Omega_\zeta$, and we deduce them from the
results in \cite{parra10a}, obtained in the laboratory frame.

In reference \cite{parra10a} we showed that in the limit $B_p/B
\ll 1$, assuming that the turbulence does not scale strongly with
$B_p/B$, the ion distribution function is given by $f_i (\bR, E,
\mu, t) = f_{Mi} (\psi(\bR), E, t) + F_{i1}^\mathrm{nc}
(\psi(\bR), \theta(\bR), E, \mu, t) + F_{i2}^\mathrm{nc}
(\psi(\bR), \theta(\bR), E, \mu, t) + F_{i2}^\mathrm{tb}
(\psi(\bR), \theta(\bR), E, \mu, t) + f_i^\mathrm{tb} (\bR, E,
\mu, t)$, where the size of these different pieces is
$F_{i1}^\mathrm{nc} \sim (B/B_p) \delta_i f_{Mi}$,
$F_{i2}^\mathrm{nc} \sim (B/B_p)^2 \delta_i^2 f_{Mi}$,
$F_{i2}^\mathrm{tb} \sim (B/B_p) (v_{ti}/qR\nu_{ii}) \delta_i^2
f_{Mi}$ and $f_i^\mathrm{tb} = f_{i1}^\mathrm{tb} +
f_{i2}^\mathrm{tb}$, with $f_{i1}^\mathrm{tb} \sim \delta_i
f_{Mi}$ and $f_{i2}^\mathrm{tb} \sim (B/B_p) \delta_i^2 f_{Mi}$.
The equations for the different pieces were obtained from the
gyrokinetic equation
\begin{equation} \label{FPgyro_original}
\frac{\partial f_i}{\partial t} + \dot{\bR} \cdot \nabla_\bR f_i +
\dot{E} \frac{\partial f_i}{\partial E} = \langle C_{ii} \{ f_i \}
\rangle_i,
\end{equation}
where the time derivative $\dot{\bR}$ is
\begin{equation}
\dot{\bR} = u^\prime \bun (\bR) + \bv_M^\prime - \frac{c}{B}
\nabla_\bR \langle \phi \rangle_i \times \bun
\end{equation}
and the time derivative $\dot{E}$ is
\begin{equation}
\dot{E} = - \frac{Ze}{M} [u^\prime \bun (\bR) + \bv_M^\prime]
\cdot \nabla_\bR \langle \phi \rangle_i.
\end{equation}
Here, $u^\prime = \pm \sqrt{2(E - \mu B)}$ is the gyrokinetic
parallel velocity in the laboratory frame, and
\begin{equation}
\bv_M^\prime = \frac{\mu}{\Omega_i} \bun \times \nabla_\bR B +
\frac{(u^\prime)^2}{\Omega_i} \bun \times (\bun \cdot \nabla_\bR
\bun)
\end{equation}
are the $\nabla B$ and curvature drifts in the laboratory frame.
Equations (19) and (20) of \cite{parra10a} for
$F_{i1}^\mathrm{nc}$ and equation (24) of \cite{parra10a} for
$F_{i2}^\mathrm{nc}$ are obtained from the long wavelength
axisymmetric contributions to \eq{FPgyro_original} of order
$\delta_i f_{Mi} v_{ti}/a$ and $(B/B_p) \delta_i^2 f_{Mi}
v_{ti}/a$, respectively. Equation (25) of \cite{parra10a} for
$F_{i2}^\mathrm{tb}$ is also a long wavelength axisymmetric
component of \eq{FPgyro_original}. In particular, it is the
contribution of order $\delta_i^2 f_{Mi} v_{ti}/a$ that when the
equation is orbit averaged does not vanish as $\nu_{ii}
\rightarrow 0$. Equation (55) of \cite{parra10a} for
$f_i^\mathrm{tb}$ is the sum of the short wavelength components of
\eq{FPgyro_original} of order $\delta_i f_{Mi} v_{ti}/a$ and
$(B/B_p) \delta_i^2 f_{Mi} v_{ti}/a$.

In this article, we write the formulation in \cite{parra10a} in
the frame rotating with velocity $\Omega_\zeta$, that is, we need
to use the new gyrokinetic variable $\varepsilon = E - I
\Omega_\zeta u^\prime/B$. Thus, the new gyrokinetic equation is
\begin{equation} \label{FPgyro_new2}
\frac{\partial f_i}{\partial t} + \dot{\bR} \cdot \nabla_\bR f_i +
\dot{\varepsilon} \frac{\partial f_i}{\partial \varepsilon} =
\langle C_{ii} \{ f_i \} \rangle_i.
\end{equation}
The time derivative of the new gyrokinetic variable $\varepsilon$
is
\begin{equation} \label{varepsilondot_aux1}
\dot{\varepsilon} = \dot{\bR} \cdot \nabla_\bR \varepsilon +
\dot{E} \frac{\partial \varepsilon}{\partial E}.
\end{equation}
In $\dot{\bR}$, using $u^\prime = u + I \Omega_\zeta /B$, with $u
= \pm \sqrt{2 (\varepsilon - \mu B + R^2 \Omega_\zeta^2/2)}$,
leads to
\begin{equation} \label{Rdot_aux}
\dot{\bR} = u \bun + \frac{I \Omega_\zeta}{B} \bun + \bv_M + \bv_C
- \frac{c}{B} \nabla_\bR \langle \phi \rangle_i \times \bun + O
\left ( \frac{B^2}{B_p^2} \delta_i^3 v_{ti} \right ),
\end{equation}
with
\begin{equation}
\bv_M = \frac{\mu}{\Omega_i} \bun \times \nabla_\bR B +
\frac{u^2}{\Omega_i} \bun \times (\bun \cdot \nabla_\bR \bun)
\end{equation}
the $\nabla B$ and curvature drifts in the rotating frame, and
$\bv_C = (2 u I \Omega_\zeta/B \Omega_i) \bun \times (\bun \cdot
\nabla_\bR \bun)$ the Coriolis drift. To obtain this expression
for $\dot{\bR}$ we have used $(u^\prime)^2 = u^2 + 2 I
\Omega_\zeta u/B + O[ (B/B_p)^2 \delta_i^2 v_{ti}^2]$ to write
$\bv_M^\prime = \bv_M + \bv_C + O[ (B^2/B_p^2) \delta_i^3 v_{ti}
]$. The usual result for the Coriolis drift $\bv_C = (2u
\Omega_\zeta/\Omega_i) \bun \times [ (\nabla_\bR R \times \zun)
\times \bun ]$ can be recovered by realizing that for $B_p/B \ll
1$, $\bun = \zun + O(B_p/B)$, $\bun \cdot \nabla_\bR \bun = -
\nabla_\bR R/R + O[(B_p/B)R^{-1}]$ and $I/B = R +
O[(B_p^2/B^2)R]$, giving
\begin{equation}
\fl \bv_C = \frac{2 u \Omega_\zeta}{\Omega_i} \bun \times
[(\nabla_\bR R \times \zun) \times \bun] = \frac{2 I u
\Omega_\zeta}{B \Omega_i} \bun \times (\bun \cdot \nabla_\bR \bun)
+ O( \delta_i^2 v_{ti}).
\end{equation}
In addition, using $I \bun/B = R \zun + \bun \times \nabla
\psi/B$, $\phi = \phi_0 + \phi_1^\mathrm{nc} + \phi_2^\mathrm{nc}
+ \phi^\mathrm{tb}$, $\langle \phi_0 \rangle_i = \phi_0
(\psi(\bR), t) + O(\delta_i^2 T_e/e)$, $\langle \phi_1^\mathrm{nc}
\rangle_i = \phi_1^\mathrm{nc} (\psi(\bR), \theta(\bR), t) +
O[(B/B_p) \delta_i^3 T_e/e]$ and $\langle \phi_2^\mathrm{nc}
\rangle_i = O[(B^2/B_p^2) \delta_i^2 v_{ti} ]$, we can simplify
equation \eq{Rdot_aux} to
\begin{eqnarray} \label{Rdot_aux2}
\fl \dot{\bR} = u \bun + R \Omega_\zeta \zun + \bv_M + \bv_C -
\frac{c}{B} \nabla_\bR \phi_1^\mathrm{nc} \times \bun -
\frac{c}{B} \nabla_\bR \langle \phi^\mathrm{tb} \rangle_i \times
\bun + O \left ( \delta_i^2 v_{ti} \right ).
\end{eqnarray}
The time derivative $\dot{\varepsilon}$ in \eq{varepsilondot_aux1}
can be written as
\begin{equation}
\dot{\varepsilon} = \dot{E} - \frac{I u^\prime}{B} \frac{\partial
\Omega_\zeta}{\partial \psi} \dot{\bR} \cdot \nabla_\bR \psi -
\Omega_\zeta \dot{\bR} \cdot \nabla_\bR \left ( \frac{I
u^\prime}{B} \right ) - \frac{I \Omega_\zeta}{B u^\prime} \dot{E}.
\end{equation}
To simplify this equation we use
\begin{eqnarray} \label{auxresult}
\fl \dot{\bR} \cdot \nabla_\bR \left ( \frac{I u^\prime}{B} \right
) = u^\prime \bun \cdot \nabla_\bR \left ( \frac{I u^\prime}{B}
\right ) + \bv^\prime_M \cdot \nabla_\bR \left ( \frac{I
u^\prime}{B} \right ) - \frac{c}{B} (\nabla_\bR \langle \phi
\rangle_i \times \bun) \cdot \nabla_\bR \left ( \frac{I
u^\prime}{B} \right ) \nonumber\\ = \frac{Ze}{Mc} \bv_M^\prime
\cdot \nabla_\bR \psi + \frac{Ze I}{M B u^\prime} \bv_M^\prime
\cdot \nabla_\bR \langle \phi \rangle_i + O \left ( \frac{B_p}{B}
\delta_i v_{ti}^2 \right ).
\end{eqnarray}
With this result, obtained by using those that follow in \eq{aux2}
and \eq{aux3}, and employing $\phi = \phi_0 + \phi_1^\mathrm{nc} +
\phi_2^\mathrm{nc} + \phi^\mathrm{tb}$, $\langle \phi_0 \rangle_i
= \phi_0 (\psi(\bR), t) + O( \delta_i^2 T_e/e)$, $\langle
\phi_1^\mathrm{nc} \rangle_i = \phi_1^\mathrm{nc} (\psi(\bR),
\theta(\bR), t) + O[(B/B_p) \delta_i^3 T_e/e]$, $\langle
\phi_2^\mathrm{nc} \rangle_i = \phi_2^\mathrm{nc} (\psi(\bR),
\theta(\bR), t) + O [ (B^2/B_p^2) \delta_i^4 T_e/e]$, $u^\prime =
u + O[(B/B_p) \delta_i v_{ti}]$, we find
\begin{eqnarray} \label{varepsilondot_aux2}
\fl \dot{\varepsilon} = - \frac{Ze}{M} [ u \bun (\bR) + \bv_M +
\bv_C] \cdot \left ( \nabla_\bR \phi_1^\mathrm{nc} + \nabla_\bR
\phi_2^\mathrm{nc} + \nabla_\bR \langle \phi^\mathrm{tb} \rangle_i
\right ) \nonumber \\ - \frac{I u}{B} \frac{\partial
\Omega_\zeta}{\partial \psi} \left ( \bv_M - \frac{c}{B}
\nabla_\bR \langle \phi^\mathrm{tb} \rangle_i \times \bun \right )
\cdot \nabla_\bR \psi + O \left ( \frac{\delta_i^2 v_{ti}^3}{a}
\right ).
\end{eqnarray}
To obtain the result in \eq{auxresult}, we have employed
$\bv_M^\prime \cdot \nabla_\bR \psi = u^\prime \bun \cdot
\nabla_\bR ( Iu^\prime/\Omega_i)$;
\begin{eqnarray} \label{aux2}
\fl - \frac{c}{B} (\nabla_\bR \langle \phi \rangle_i \times \bun)
\cdot \nabla_\bR \left ( \frac{I u^\prime}{B} \right ) \nonumber\\
= \frac{ZeI}{MB u^\prime} \left [ \frac{\mu }{\Omega_i} \bun
\times \nabla_\bR B - \frac{(u^\prime)^2}{\Omega_i} \bun \times
\nabla_\bR \ln \left ( \frac{I}{B} \right ) \right ] \cdot
\nabla_\bR \langle \phi \rangle_i \nonumber\\ = \frac{ZeI}{MB
u^\prime} \left [ \frac{\mu }{\Omega_i} \bun \times \nabla_\bR B +
\frac{(u^\prime)^2}{\Omega_i} \bun \times ( \bun \cdot \nabla_\bR
\bun) \right ] \cdot \nabla_\bR \langle \phi \rangle_i + O \left (
\frac{B_p}{B} \delta_i v_{ti}^2 \right ) \nonumber\\ =
\frac{ZeI}{MB u^\prime} \bv_M^\prime \cdot \nabla_\bR \langle \phi
\rangle_i + O \left ( \frac{B_p}{B} \delta_i v_{ti}^2 \right ),
\end{eqnarray}
where we have used $I/B = R + O[(B_p^2/B^2)R]$ and $\bun \cdot
\nabla_\bR \bun = - \nabla_\bR \ln R + O[(B_p/B)R^{-1}]$; and
\begin{equation} \label{aux3}
\fl \bv^\prime_M \cdot \nabla_\bR \left ( \frac{I u^\prime}{B}
\right ) = \frac{u^\prime}{\Omega_i} \left [ \nabla_\bR \times (
u^\prime \bun) - u^\prime \bun \bun \cdot \nabla_\bR \times \bun
\right ] \cdot \nabla_\bR \left ( \frac{I u^\prime}{B} \right )
\sim \frac{B_p^2}{B^2} \delta_i v_{ti}^2,
\end{equation}
where we have used $\bun \cdot \nabla_\bR \times \bun \sim
(B_p/B)a^{-1}$, $\bun \cdot \nabla_\bR (Iu^\prime/B) \sim (a/R)(R
v_{ti}/qR) \sim (B_p/B) v_{ti}$ and $\nabla_\bR \times (u^\prime
\bun) \cdot \nabla_\bR ( I u^\prime/B) = \nabla_\bR \cdot [
u^\prime \bun \times \nabla_\bR ( I u^\prime/B) ] = \nabla_\bR
\cdot [ (I u^\prime/B) \nabla_\bR \zeta \times \nabla_\bR (I
u^\prime/B) ] + \nabla_\bR \cdot [ (u^\prime/B) (\nabla \zeta
\times \nabla \psi) \times \nabla_\bR ( I u^\prime/B) ] =
\nabla_\bR \cdot \{ \nabla_\bR \zeta \times\nabla_\bR [I^2
(u^\prime)^2/2B^2] \} -
\partial_\zeta [ (u^\prime/R^2 B) \nabla \psi
\cdot \nabla_\bR ( I u^\prime/B ) ] = 0$.

With equations \eq{FPgyro_new2}, \eq{Rdot_aux2} and
\eq{varepsilondot_aux2}, we can now easily obtain equations
\eq{eqHi1nc}, \eq{eqfitb} and \eq{eqF2} for $H_{i1}^\mathrm{nc}$,
$f_i^\mathrm{tb}$, $H_{i2}^\mathrm{nc}$ and $H_{i2}^\mathrm{tb}$.
To obtain \eq{eqHi1nc}, we take the long wavelength axisymmetric
contribution to \eq{FPgyro_new2} to order $\delta_i f_{Mi}
v_{ti}/a$, giving
\begin{eqnarray} \label{eqHi1nc_aux}
\fl u \bun \cdot \nabla_\bR H_{i1}^\mathrm{nc} + \bv_M \cdot
\nabla_\bR f_{Mi} - \frac{Ze}{M} \frac{\partial f_{Mi}}{\partial
\varepsilon} u \bun \cdot \nabla_\bR \phi_1^\mathrm{nc} =
C_{ii}^{(\ell)} \{ H_{i1}^\mathrm{nc} \}.
\end{eqnarray}
This equation differs from equations (19) and (20) of
\cite{parra10a}, and gives a function $H_{i1}^\mathrm{nc}$
different from the function $F_{i1}^\mathrm{nc}$ defined in
\cite{parra10a}. The reason is that $f_{Mi} (\psi(\bR),
\varepsilon) + H_{i1}^\mathrm{nc} + H_{i2}^\mathrm{nc}$ must be
equal to the function $f_{Mi} (\psi(\bR), E) + F_{i1}^\mathrm{nc}
+ F_{i2}^\mathrm{nc}$ defined in \cite{parra10a} to the order of
interest, but how the terms of first and second order in
$\delta_i$ are assigned to one or the other piece differs
depending on the frame. For this reason, we have changed the name
of the functions. The final result in \eq{eqHi1nc} is obtained
from \eq{eqHi1nc_aux} by using $\bv_M \cdot \nabla_\bR \psi = u
\bun \cdot \nabla_\bR ( I u/\Omega_i)$ for $u = \pm
\sqrt{2(\varepsilon - \mu B + R^2 \Omega_\zeta^2/2)} \simeq \pm
\sqrt{2(\varepsilon - \mu B)}$.

Equation \eq{eqfitb} is the sum of the short wavelength
contributions to \eq{FPgyro_new2} of order $\delta_i f_{Mi}
v_{ti}/a$ and $(B/B_p) \delta_i^2 f_{Mi} v_{ti}/a$. The equation
is straightforward if we apply the same methodology as in
\cite{parra10a}.

Equation \eq{eqF2} is found from the long wavelength axisymmetric
components of \eq{FPgyro_new2} to order $\delta_i^2 f_{Mi}
v_{ti}/a$. Note that to this order we have the time derivative
$\partial_t f_{Mi}$ \cite{parra10a}. Using $\partial_t f_{Mi} = [
n_i^{-1}\partial_t n_i + (M\varepsilon/T_i - 3/2)
T_i^{-1}\partial_t T_i ] f_{Mi}$ and realizing that $\partial_t
n_i = \langle \int d^3v\, \mathcal{S}^\mathrm{tb} \rangle_\psi$
and $(3/2)\partial_t (n_i T_i) = \langle \int d^3v\,
\mathcal{S}^\mathrm{tb} M \varepsilon \rangle_\psi + \langle \int
d^3v\, \mathcal{S}^\mathrm{nc} M \varepsilon \rangle_\psi$, we
find the final form in \eq{eqF2}, \eq{Snc} and \eq{Stb}. Here the
integral $\langle \int d^3v\, \mathcal{S}^\mathrm{tb} M
\varepsilon \rangle_\psi$ gives both the divergence of the
turbulent radial energy transport and the turbulent heating.
Similarly, $\langle \int d^3v\, \mathcal{S}^\mathrm{nc} M
\varepsilon \rangle_\psi$ gives the divergence of the neoclassical
radial flux of energy. The equations for $H_{i2}^\mathrm{nc}$ and
$H_{i2}^\mathrm{tb}$ are obtained in the same way as equations
(24) and (25) in \cite{parra10a}, i.e., the equation for
$H_{i2}^\mathrm{nc}$ is the axisymmetric long wavelength component
of \eq{FPgyro_new2} of order $(B/B_p) \delta_i^2 f_{Mi} v_{ti}/a$,
and the equation for $H_{i2}^\mathrm{tb}$ is the axisymmetric long
wavelength component of order $\delta_i^2 f_{Mi} v_{ti}/a$ that
when it is orbit averaged does not vanish as $\nu_{ii} \rightarrow
0$.

The equations \eq{eqHe1nc} and \eq{eqfetb} for the electron
distribution function in the rotating frame are derived in the
same way as the equations for the ion distribution function. The
only differences are that the Coriolis drift $\bv_C$ and the term
in \eq{varepsilondot_aux2} that is proportional to $\partial_\psi
\Omega_\zeta$ are small by $\sqrt{m/M}$ and hence negligible, and
that we include the electric field $\bE^A$ driven by the
transformer, leading to a modified time derivative for the energy
\begin{equation}
\fl \dot{\varepsilon} = - \frac{e}{m} u \bun \cdot \bE^A +
\frac{e}{m} [ u \bun (\bR) + \bv_M] \cdot \left (\nabla_\bR
\phi_1^\mathrm{nc} + \nabla_\bR \langle \phi^\mathrm{tb} \rangle_e
\right ).
\end{equation}

Finally, the equations for the different pieces of the
electrostatic potential \eq{eqphi1nc} and \eq{eqphitb} are easily
deduced from the results in \cite{parra10a} by realizing that
moving to a rotating reference frame does not modify the
quasineutrality equation.

%%%%%%%%%%%%%%%%%%%%%%%%%%%%%%%%%%%%%%%%%%%%%%%%%%%%%%%%%%%%%%%%%%%%%
\section{Derivation of equation \eq{pi1}} \alabel{app_pi}
%%%%%%%%%%%%%%%%%%%%%%%%%%%%%%%%%%%%%%%%%%%%%%%%%%%%%%%%%%%%%%%%%%%%%

In this Appendix we derive equation~\eq{pi1} with a procedure
different from the one employed in \cite{parra10a}. This new
derivation shows the connection with calculations that split the
off-diagonal components of the viscosity into gyroviscosity and
perpendicular viscosity \cite{simakov07}. The derivation presented
here and the one in \cite{parra10a} lead to identical results (as
they should), but we believe that this new approach emphasizes the
advantages of the formula in \eq{pi1}.

We begin by using $R \zun = I\bun/B - B^{-1} \bun \times \nabla
\psi$ to write
\begin{eqnarray}
\fl \langle \langle R \zun \cdot \matrixtop{\mathbf{P}}_i \cdot
\nabla \psi \rangle_\psi \rangle_\mathrm{T} = \left \langle \left
\langle \frac{M}{B} \int d^3v^\prime\, f_i \bv_\bot^\prime \cdot
\nabla \psi \left [ I v_{||}^\prime - (\bv^\prime \times \bun)
\cdot \nabla \psi \right ] \right \rangle_\psi \right
\rangle_\mathrm{T} \nonumber\\= \left \langle \left \langle
\frac{M I}{B} \int d^3v^\prime\, f_i v_{||}^\prime \bv_\bot^\prime
\cdot \nabla \psi \right \rangle_\psi \right \rangle_\mathrm{T}
\nonumber\\- \left \langle \left \langle \frac{M}{2B} \int
d^3v^\prime\, f_i \nabla \psi \cdot \left [ \bv_\bot^\prime
\bv_\bot^\prime - (\bv^\prime \times \bun) (\bv^\prime \times
\bun) \right ] \cdot ( \bun \times \nabla \psi ) \right
\rangle_\psi \right \rangle_\mathrm{T}.
\end{eqnarray}
Since the transport of toroidal angular momentum needs to be known
to order $(B/B_p) \delta_i^3 p_i R |\nabla \psi|$, evaluating it
directly from this equation would require knowing $f_i$ to order
$(B/B_p) \delta_i^3 f_{Mi}$, and it is easy to see from our
decomposition of the ion distribution function given in
Section~\ref{sub_f} that we cannot calculate the ion distribution
function to that order. To circumvent this problem, we use exact
moments of the Fokker-Planck equation to write the two integrals
that appear in the transport of momentum as
\begin{eqnarray} \label{pi_parperp}
\fl M \int d^3v^\prime\, f_i v_{||}^\prime \bv_\bot^\prime = -
\frac{M}{\Omega_i} \frac{\partial}{\partial t} \left [ \int
d^3v^\prime\, f_i v_{||}^\prime (\bv^\prime \times \bun) \right ]
+ \frac{M}{\Omega_i} \bun \times \left [ \nabla \cdot \left ( \int
d^3v^\prime\, f_i \bv^\prime \bv^\prime \bv^\prime \right ) \right
] \cdot \bun \nonumber\\- \frac{Mc}{B} \left [ \bun \cdot \nabla
\phi \int d^3v^\prime\, f_i \bv^\prime \times \bun + (\nabla \phi
\times \bun ) \int d^3v^\prime\, f_i v_{||}^\prime \right ]
\nonumber\\+ \frac{M}{\Omega_i} \int d^3v^\prime\, C_{ii} \{ f_i
\} v_{||}^\prime (\bv^\prime \times \bun),
\end{eqnarray}
where we have used the $v_{||}^\prime (\bv^\prime \times \bun)$
moment of the Fokker-Planck equation, and
\begin{eqnarray} \label{pi_perpperp}
\fl \frac{M}{2} \int d^3v^\prime\, f_i \left [ \bv_\bot^\prime
\bv_\bot^\prime - (\bv^\prime \times \bun) (\bv^\prime \times
\bun) \right ] \nonumber\\= - \frac{M}{4\Omega_i}
\frac{\partial}{\partial t} \left [ \int d^3v^\prime\, f_i \left (
\bv_\bot^\prime (\bv^\prime \times \bun) + (\bv^\prime \times
\bun) \bv_\bot^\prime \right ) \right ] \nonumber\\+
\frac{M}{4\Omega_i} \bun \times \left [ \nabla \cdot \left ( \int
d^3v^\prime\, f_i \bv^\prime \bv^\prime \bv^\prime \right ) \right
] \cdot \left ( \matI - \bun \bun \right ) \nonumber\\-
\frac{M}{4\Omega_i} \left ( \matI - \bun \bun \right ) \cdot \left
[ \nabla \cdot \left ( \int d^3v^\prime\, f_i \bv^\prime
\bv^\prime \bv^\prime \right ) \right ] \times \bun \nonumber\\-
\frac{Mc}{4B} \int d^3v^\prime\, f_i \left [ (\nabla \phi \times
\bun) \bv_\bot^\prime + \bv_\bot^\prime (\nabla \phi \times \bun)
\right ] \nonumber\\- \frac{Mc}{4B} \int d^3v^\prime\, f_i \left [
\nabla_\bot \phi (\bv^\prime \times \bun) + (\bv^\prime \times
\bun) \nabla_\bot \phi \right ] \nonumber\\+ \frac{M}{4\Omega_i}
\int d^3v^\prime\, C_{ii} \{ f_i \} \left [ \bv_\bot^\prime
(\bv^\prime \times \bun) + (\bv^\prime \times \bun)
\bv_\bot^\prime \right ],
\end{eqnarray}
where we have used the $\bv_\bot^\prime (\bv^\prime \times \bun) +
(\bv^\prime \times \bun) \bv_\bot^\prime$ moment. These are the
standard expressions used to calculate the perpendicular-parallel
and perpendicular-perpendicular components of gyroviscosity and
perpendicular viscosity \cite{simakov07}.

Upon coarse grain averaging, the terms with time derivatives in
\eq{pi_parperp} and \eq{pi_perpperp} are smaller than $\delta_i^3
p_i$ because $\partial_t$ becomes of order $\delta_i^2 v_{ti}/a$
and the lowest order distribution function is a Maxwellian, making
the velocity integrals over $v_{||}^\prime (\bv^\prime \times
\bun)$ and $\bv^\prime_\bot ( \bv^\prime \times \bun) + (
\bv^\prime \times \bun) \bv_\bot^\prime$ vanish. As a result, the
time derivative terms are negligible, giving
\begin{eqnarray} \label{pi_parperp2}
\fl \left \langle \left \langle \frac{MI}{B} \int d^3v^\prime\,
f_i v_{||}^\prime \bv_\bot^\prime \cdot \nabla \psi \right
\rangle_\psi \right \rangle_\mathrm{T} \nonumber\\= - \left
\langle \frac{M I}{B \Omega_i} (\bun \times \nabla \psi) \cdot
\left [ \nabla \cdot \left ( \int d^3v^\prime\, \langle f_i
\rangle_\mathrm{T} \bv^\prime \bv^\prime \bv^\prime \right )
\right ] \cdot \bun \right \rangle_\psi \nonumber\\- \left \langle
\left \langle \frac{McI}{B^2} \bun \cdot \nabla \phi \int
d^3v^\prime\, f_i (\bv^\prime \times \bun ) \cdot \nabla \psi
\right \rangle_\psi \right \rangle_\mathrm{T} \nonumber\\- \left
\langle \left \langle \frac{McI}{B^2} (\nabla \phi \times \bun)
\cdot \nabla \psi \int d^3v^\prime\, f_i v_{||}^\prime \right
\rangle_\psi \right \rangle_\mathrm{T} \nonumber\\+ \left \langle
\frac{MI}{B\Omega_i} \int d^3v^\prime\, \left \langle C_{ii} \{
f_i \} \right \rangle_\mathrm{T} v_{||}^\prime (\bv^\prime \times
\bun ) \cdot \nabla \psi \right \rangle_\psi
\end{eqnarray}
and
\begin{eqnarray} \label{pi_perpperp2}
\fl \left \langle \left \langle \frac{M}{2B} \int d^3v^\prime\,
f_i \nabla \psi \cdot \left [ \bv_\bot^\prime \bv_\bot^\prime -
(\bv^\prime \times \bun) (\bv^\prime \times \bun) \right ] \cdot (
\bun \times \nabla \psi ) \right \rangle_\psi \right
\rangle_\mathrm{T} \nonumber\\= \left \langle \frac{M}{4B\Omega_i}
\nabla \psi \cdot \left [ \nabla \cdot \left ( \int d^3v^\prime\,
\left \langle f_i \right \rangle_\mathrm{T} \bv^\prime \bv^\prime
\bv^\prime \right ) \right ] \cdot \nabla \psi \right \rangle_\psi
\nonumber\\- \left \langle \frac{M}{4B\Omega_i} (\bun \times
\nabla \psi) \cdot \left [ \nabla \cdot \left ( \int d^3v^\prime\,
\left \langle f_i \right \rangle_\mathrm{T} \bv^\prime \bv^\prime
\bv^\prime \right ) \right ] \cdot ( \bun \times \nabla \psi )
\right \rangle_\psi \nonumber\\+ \left \langle \left \langle
\frac{Mc}{2B^2} \int d^3v^\prime\, f_i ( \bv^\prime \cdot \nabla
\psi) (\nabla \phi \cdot \nabla \psi) \right \rangle_\psi \right
\rangle_\mathrm{T}\nonumber\\- \left \langle \left \langle
\frac{Mc}{2B^2} \int d^3v^\prime\, f_i [ (\bv^\prime \times \bun)
\cdot \nabla \psi ] [ (\nabla \phi \times \bun) \cdot \nabla \psi
] \right \rangle_\psi \right \rangle_\mathrm{T} \nonumber\\- \left
\langle \frac{M}{4B\Omega_i} \int d^3v^\prime\, \left \langle
C_{ii} \{ f_i \} \right \rangle_\mathrm{T} \left [ ( \bv^\prime
\cdot \nabla \psi)^2  - \left ( (\bv^\prime \times \bun) \cdot
\nabla \psi \right )^2 \right ] \right \rangle_\psi.
\end{eqnarray}
We could use these two equations to evaluate $\langle \langle R
\zun \cdot \matrixtop{\mathbf{P}}_i \cdot \nabla \psi \rangle_\psi
\rangle_\mathrm{T}$ instead of the form in \eq{pi1}. If we did so,
we would still need to evaluate the first term on the right side
of \eq{pi_parperp2} and the first and second terms on the right
side of \eq{pi_perpperp2} to order $\delta_i^2 \Delta_{ud} p_i R
|\nabla \psi|$ for an up-down asymmetric tokamak with $\Delta_{ud}
\gg (B/B_p) \delta_i$, and to order $(B/B_p) \delta_i^3 p_i R
|\nabla \psi|$ for an up-down symmetric tokamak with $\Delta_{ud}
\ssim (B/B_p) \delta_i$. In the up-down asymmetric case, the
dominant contribution to the first term on the right side of
\eq{pi_parperp2} and the first and second terms on the right side
of \eq{pi_perpperp2} is due to $H_{i1}^\mathrm{nc}$,
\begin{eqnarray}
\fl - \left \langle \frac{M I}{B \Omega_i} (\bun \times \nabla
\psi) \cdot \left [ \nabla \cdot \left ( \int d^3v^\prime\,
H_{i1,0}^\mathrm{nc} \bv^\prime \bv^\prime \bv^\prime \right )
\right ] \cdot \bun \right \rangle_\psi \nonumber\\\simeq \left
\langle \frac{M I}{B \Omega_i} \bun \cdot \nabla \bun \cdot (\bun
\times \nabla \psi) \int d^3v\, H_{i1}^\mathrm{nc} v_{||}
(v_\bot^2 - v_{||}^2) \right \rangle_\psi \nonumber\\- \left
\langle \frac{M I}{2B \Omega_i} (\bun \times \nabla \psi) \cdot
\nabla \left ( \int d^3v\, H_{i1}^\mathrm{nc} v_{||} v_\bot^2
\right ) \right \rangle_\psi \nonumber\\= \frac{M^2 c}{2 Z e}
\left \langle \int d^3v\, H_{i1}^\mathrm{nc} v_{||} \left [
v_{||}^2 \bun \cdot \nabla \left ( \frac{I^2}{B^2} \right ) -
\frac{I^2}{B^2} v_\bot^2 \bun \cdot \nabla \ln B \right ] \right
\rangle_\psi.
\end{eqnarray}
This term is of order $(B/B_p) \Delta_{ud} \delta_i^2 p_i R
|\nabla \psi|$, suggesting that the transport due to the
neoclassical piece $H_{i1}^\mathrm{nc}$ does not scale with
collisionality and that it is larger than the turbulent piece
$\Pi_{-1}^\mathrm{tb}$ by a factor of $B/B_p \gg 1$ (see
Table~\ref{table_Pi}). In fact, this contribution cancels to this
order with other terms in \eq{pi_parperp2} and \eq{pi_perpperp2},
as we will show in equations \eq{auxgrad}, \eq{auxphi}, \eq{auxC},
\eq{auxppar} and \eq{auxpperp} below. The final result of the
cancellation is that the neoclassical pieces of the distribution
function give a contribution of order $(B/B_p)(qR\nu_{ii}/v_{ti})
\Delta_{ud} \delta_i^2 p_i R |\nabla \psi|$ for $\Delta_{ud} \gg
(B/B_p) \delta_i$.

In the up-down symmetric case, similar problems appear. Obtaining
the transport of toroidal angular momentum to order $(B/B_p)
\delta_i^3 p_i R |\nabla \psi|$ requires calculating the long
wavelength piece of the ion distribution function $\langle f_i
\rangle_\mathrm{T}$ to order $(B/B_p) \delta_i^2 f_{Mi}$ for the
first term on the right side of \eq{pi_parperp2} and to order
$(B/B_p)^2 \delta_i^2 f_{Mi}$ for the first and second terms on
the right side of \eq{pi_perpperp2}. In section~\ref{sub_f} we
show how to calculate $H_{i2}^\mathrm{nc} \sim (B/B_p)^2
\delta_i^2 f_{Mi}$ and $H_{i2}^\mathrm{tb} \sim (B/B_p) (v_{ti}/qR
\nu_{ii}) \delta_i^2 f_{Mi}$, so in principle, it is possible to
evaluate these terms to the appropriate order. Note, however, that
equations \eq{pi_parperp2} and \eq{pi_perpperp2} suggest
misleading scalings for the transport of momentum. For example,
using $H_{i2}^\mathrm{nc} \sim (B/B_p)^2 \delta_i^2 f_{Mi}$ in the
first and second terms on the right side of \eq{pi_perpperp2}, we
obtain that a purely neoclassical piece can give momentum
transport of order $(B/B_p) \delta_i^3 p_i R |\nabla \psi|$, that
is, transport that does not scale with collisionality. Similarly,
integrating over $H_{i2}^\mathrm{tb} \sim (B/B_p)
(v_{ti}/qR\nu_{ii}) \delta_i^2 f_{Mi}$ in the first term on the
right side of \eq{pi_parperp2} gives momentum transport of order
$(B/B_p) (v_{ti}/qR\nu_{ii}) \delta_i^3 p_i R |\nabla \psi|$, i.e.
it scales inversely with collisionality. In fact, these
contributions to the transport of momentum vanish by themselves or
when combined with other terms in \eq{pi_parperp2} and
\eq{pi_perpperp2} as we will show in \eq{auxgrad}, \eq{auxphi},
\eq{auxC}, \eq{auxppar} and \eq{auxpperp}. In equation \eq{pi1}
these cancellations have already been taken into account. There
are advantages to this. For example, if we decide to use
\eq{pi_parperp2} and \eq{pi_perpperp2} instead of \eq{pi1},
$H_{i2}^\mathrm{tb} \sim (B/B_p) (v_{ti}/qR\nu_{ii}) \delta_i^2
f_{Mi}$ must be calculated to order $(B/B_p) \delta_i^2 f_{Mi} <
(B/B_p) (v_{ti}/qR\nu_{ii}) \delta_i^2 f_{Mi}$, that is, to more
precision than is necessary. Equation \eq{pi1}, on the other hand,
makes explicit that the first term in the right side of
\eq{pi_parperp2} vanishes when the lowest order piece of
$H_{i2}^\mathrm{nc} \sim (B/B_p) (v_{ti}/qR\nu_{ii}) \delta_i^2
f_{Mi}$ is integrated over and combined with other terms in
\eq{pi_parperp2} and \eq{pi_perpperp2}.

In what follows we show how to obtain \eq{pi1} from
\eq{pi_parperp2} and \eq{pi_perpperp2}. Using $\nabla \psi \nabla
\psi = |\nabla \psi|^2 ( \matI - \bun \bun ) - (\bun \times \nabla
\psi) (\bun \times \nabla \psi)$, $\bun \times \nabla \psi = I\bun
- RB \zun$ and $\nabla (R \zun) = (\nabla R) \zun - \zun (\nabla
R)$, we obtain
\begin{eqnarray} \label{auxgrad}
\fl - \left \langle \frac{M I}{B \Omega_i} (\bun \times \nabla
\psi) \cdot \left [ \nabla \cdot \left ( \int d^3v^\prime\,
\langle f_i \rangle_\mathrm{T} \bv^\prime \bv^\prime \bv^\prime
\right ) \right ] \cdot \bun \right \rangle_\psi \nonumber\\-
\left \langle \frac{M}{4B\Omega_i} \nabla \psi \cdot \left [
\nabla \cdot \left ( \int d^3v^\prime\, \left \langle f_i \right
\rangle_\mathrm{T} \bv^\prime \bv^\prime \bv^\prime \right )
\right ] \cdot \nabla \psi \right \rangle_\psi \nonumber\\+ \left
\langle \frac{M}{4B\Omega_i} (\bun \times \nabla \psi) \cdot \left
[ \nabla \cdot \left ( \int d^3v^\prime\, \left \langle f_i \right
\rangle_\mathrm{T} \bv^\prime \bv^\prime \bv^\prime \right )
\right ] \cdot ( \bun \times \nabla \psi ) \right \rangle_\psi
\nonumber\\ = \frac{M^2c}{2Ze} \frac{1}{V^\prime}
\frac{\partial}{\partial \psi} V^\prime \left \langle R^2 \int
d^3v^\prime\, \langle f_i \rangle_\mathrm{T} (\bv^\prime \cdot
\zun)^2 \bv^\prime \cdot \nabla \psi \right \rangle_\psi
\nonumber\\- \left \langle \frac{M I^2}{2 B \Omega_i} \bun \cdot
\left [ \nabla \cdot \left ( \int d^3v^\prime\, \langle f_i
\rangle_\mathrm{T} \bv^\prime \bv^\prime \bv^\prime \right )
\right ] \cdot \bun \right \rangle_\psi \nonumber\\- \left \langle
\frac{M |\nabla \psi|^2}{4 B \Omega_i} \left ( \matI - \bun \bun
\right ): \left [ \nabla \cdot \left ( \int d^3v^\prime\, \langle
f_i \rangle_\mathrm{T} \bv^\prime \bv^\prime \bv^\prime \right )
\right ] \right \rangle_\psi,
\end{eqnarray}
\begin{eqnarray} \label{auxphi}
\fl - \left \langle \left \langle \frac{McI}{B^2} \bun \cdot
\nabla \phi \int d^3v^\prime\, f_i (\bv^\prime \times \bun ) \cdot
\nabla \psi \right \rangle_\psi \right \rangle_\mathrm{T}
\nonumber\\- \left \langle \left \langle \frac{McI}{B^2} (\nabla
\phi \times \bun) \cdot \nabla \psi \int d^3v^\prime\, f_i
v_{||}^\prime \right \rangle_\psi \right \rangle_\mathrm{T}
\nonumber\\- \left \langle \left \langle \frac{Mc}{2B^2} \int
d^3v^\prime\, f_i ( \bv^\prime \cdot \nabla \psi) (\nabla \phi
\cdot \nabla \psi) \right \rangle_\psi \right
\rangle_\mathrm{T}\nonumber\\+ \left \langle \left \langle
\frac{Mc}{2B^2} \int d^3v^\prime\, f_i [ (\bv^\prime \times \bun)
\cdot \nabla \psi ] [ (\nabla \phi \times \bun) \cdot \nabla \psi
] \right \rangle_\psi \right \rangle_\mathrm{T} \nonumber\\ =
Mc\left \langle \left \langle R^2 \zun \cdot \nabla \phi \int
d^3v^\prime\, f_i (\bv^\prime \cdot \zun) \right \rangle_\psi
\right \rangle_\mathrm{T} \nonumber\\- \left \langle \left \langle
\frac{McI^2}{B^2} \bun \cdot \nabla \phi \int d^3v^\prime\, f_i
v_{||}^\prime \right \rangle_\psi \right \rangle_\mathrm{T}
\nonumber\\- \left \langle \left \langle \frac{Mc|\nabla
\psi|^2}{2B^2} \int d^3v^\prime\, f_i (\bv_\bot^\prime \cdot
\nabla \phi ) \right \rangle_\psi \right \rangle_\mathrm{T}
\end{eqnarray}
and
\begin{eqnarray} \label{auxC}
\fl \left \langle \frac{MI}{B\Omega_i} \int d^3v^\prime\, \left
\langle C_{ii} \{ f_i \} \right \rangle_\mathrm{T} v_{||}^\prime
(\bv^\prime \times \bun ) \cdot \nabla \psi \right \rangle_\psi
\nonumber\\+ \left \langle \frac{M}{4B\Omega_i} \int d^3v^\prime\,
\left \langle C_{ii} \{ f_i \} \right \rangle_\mathrm{T} \left [ (
\bv^\prime \cdot \nabla \psi)^2  - \left ( (\bv^\prime \times
\bun) \cdot \nabla \psi \right )^2 \right ] \right \rangle_\psi
\nonumber\\= - \frac{M^2c}{2Ze} \left \langle R^2 \int
d^3v^\prime\, \left \langle C_{ii} \{ f_i \} \right
\rangle_\mathrm{T} (\bv^\prime \cdot \zun )^2 \right \rangle_\psi
\nonumber\\+ \left \langle \frac{MI^2}{2B\Omega_i} \int
d^3v^\prime\, \left \langle C_{ii} \{ f_i \} \right
\rangle_\mathrm{T} (v_{||}^\prime)^2 \right \rangle_\psi
\nonumber\\+ \left \langle \frac{M|\nabla \psi|^2}{4B\Omega_i}
\int d^3v^\prime\, \left \langle C_{ii} \{ f_i \} \right
\rangle_\mathrm{T} \left ( v_\bot^\prime \right )^2 \right
\rangle_\psi.
\end{eqnarray}
Furthermore, we will use the $(M I^2/2B \Omega_i)
(v_{||}^\prime)^2$ and $(M |\nabla \psi|^2/4B \Omega_i)
(v_\bot^\prime)^2$ moments of the Fokker-Planck equation to remove
most of the $\int d^3v^\prime\, \langle f_i \rangle_\mathrm{T}
\bv^\prime \bv^\prime \bv^\prime$ moments,
\begin{eqnarray} \label{auxppar}
\fl - \left \langle \frac{M I^2}{2 B \Omega_i} \bun \cdot \left [
\nabla \cdot \left ( \int d^3v^\prime\, \langle f_i
\rangle_\mathrm{T} \bv^\prime \bv^\prime \bv^\prime \right )
\right ] \cdot \bun \right \rangle_\psi - \left \langle \left
\langle \frac{McI^2}{B^2} \bun \cdot \nabla \phi \int
d^3v^\prime\, f_i v_{||}^\prime \right \rangle_\psi \right
\rangle_\mathrm{T} \nonumber\\+ \left \langle
\frac{MI^2}{2B\Omega_i} \int d^3v^\prime\, \left \langle C_{ii} \{
f_i \} \right \rangle_\mathrm{T} (v_{||}^\prime)^2 \right
\rangle_\psi \nonumber\\ = \frac{\partial}{\partial t} \left
\langle \frac{MI^2}{2B\Omega_i} \int d^3v^\prime\, \left \langle
f_i \right \rangle_\mathrm{T} (v_{||}^\prime)^2 \right
\rangle_\psi = \left \langle \frac{I^2}{2B\Omega_i} \right
\rangle_\psi \frac{\partial p_i}{\partial t}
\end{eqnarray}
and
\begin{eqnarray} \label{auxpperp}
\fl - \left \langle \frac{M |\nabla \psi|^2}{4 B \Omega_i} \left (
\matI - \bun \bun \right ) : \left [ \nabla \cdot \left ( \int
d^3v^\prime\, \langle f_i \rangle_\mathrm{T} \bv^\prime \bv^\prime
\bv^\prime \right ) \right ] \right \rangle_\psi \nonumber\\-
\left \langle \left \langle \frac{Mc|\nabla \psi|^2}{2B^2} \int
d^3v^\prime\, f_i \bv_\bot^\prime \cdot \nabla \phi \right
\rangle_\psi \right \rangle_\mathrm{T} \nonumber\\+ \left \langle
\frac{M|\nabla \psi|^2}{4B\Omega_i} \int d^3v^\prime\, \left
\langle C_{ii} \{ f_i \} \right \rangle_\mathrm{T}
(v_\bot^\prime)^2 \right \rangle_\psi \nonumber\\ =
\frac{\partial}{\partial t} \left \langle \frac{M|\nabla
\psi|^2}{4B\Omega_i} \int d^3v^\prime\, \left \langle f_i \right
\rangle_\mathrm{T} (v_\bot^\prime)^2 \right \rangle_\psi = \left
\langle \frac{|\nabla \psi|^2}{2B\Omega_i} \right \rangle_\psi
\frac{\partial p_i}{\partial t}.
\end{eqnarray}
Combining \eq{auxgrad}, \eq{auxphi} and \eq{auxC}, and then using
\eq{auxppar} and \eq{auxpperp}, we finally obtain
\begin{eqnarray}
\fl \langle \langle R\zun \cdot \matrixtop{\mathbf{P}}_i \cdot
\nabla \psi \rangle_\psi \rangle_\mathrm{T} = \frac{Mc}{2Ze}
\langle R^2 \rangle_\psi \frac{\partial p_i}{\partial t} +
\frac{M^2c}{2Ze} \frac{1}{V^\prime} \frac{\partial}{\partial \psi}
V^\prime \left \langle R^2 \int d^3v^\prime\, \langle f_i
\rangle_\mathrm{T} (\bv^\prime \cdot \zun)^2 \bv^\prime \cdot
\nabla \psi \right \rangle_\psi \nonumber\\+ Mc\left \langle \left
\langle R^2 \zun \cdot \nabla \phi \int d^3v^\prime\, f_i
(\bv^\prime \cdot \zun) \right \rangle_\psi \right
\rangle_\mathrm{T} \nonumber\\- \frac{M^2c}{2Ze} \left \langle R^2
\int d^3v^\prime\, \left \langle C_{ii} \{ f_i \} \right
\rangle_\mathrm{T} (\bv^\prime \cdot \zun )^2 \right \rangle_\psi.
\end{eqnarray}
To obtain the final result in \eq{pi1}, we just need to rewrite
the integral $\langle R^2 \int d^3v^\prime\, \langle f_i
\rangle_\mathrm{T} (\bv^\prime \cdot \zun)^2 \bv^\prime \cdot
\nabla \psi \rangle_\psi$ using the Fokker-Planck equation as is
done in \cite{parra10a}.

Finally, note that in writing \eq{pi1} we have used the
expressions \eq{pi_parperp} and \eq{pi_perpperp} that give the
gyroviscosity and perpendicular viscosity, and \eq{auxppar} and
\eq{auxpperp} that are equations for the parallel and
perpendicular pressure. It is necessary to combine all these
equations to explicitly show the cancellations mentioned below
equation \eq{pi_perpperp2}.


\section*{References}
\begin{thebibliography}{10}

\bibitem{rice07}
Rice~J~E et al 2007
\newblock {\em Nucl. Fusion} {\bf 47} 1618

\bibitem{nave10}
Nave~M~F~F et al 2010
\newblock {\em Phys. Rev. Lett.} {\bf 105} 105005

\bibitem{ikeda07}
Ikeda~K et al 2007
\newblock {\em Nucl. Fusion} {\bf 47} E01.

\bibitem{peeters07}
Peeters~A~G, Angioni~C and Strinzi~D 2007
\newblock \PRL {\bf 98} 265003

\bibitem{waltz07}
Waltz~R~E, Staebler~G~M, Candy~J and Hinton~F~L 2007
\newblock {\em Phys. Plasmas} {\bf 14} 122507 \\
Waltz~R~E, Staebler~G~M, Candy~J and Hinton~F~L 2009 {\bf 16}
079902

\bibitem{roach09}
Roach~C~M \emph{et al} 2009
\newblock \PPCF {\bf 51} 124020

\bibitem{camenen09}
Camenen~Y, Peeters~A~G, Angioni~C, Casson~F~J, Hornsby~W~A,
Snodin~A~P and Strintzi~D 2009
\newblock \PRL {\bf 102} 125001

\bibitem{camenen09b}
Camenen~Y, Peeters~A~G, Angioni~C, Casson~F~J, Hornsby~W~A,
Snodin~A~P and Strintzi~D 2009
\newblock {\em Phys. Plasmas} {\bf 16} 062501

\bibitem{casson09}
Casson~F~J, Peeters~A~G, Camenen~Y, Angioni~C, Hornsby~W~A,
Snodin~A~P, Strintzi~D and Szepesi~G 2009
\newblock {\em Phys. Plasmas} {\bf 16} 092303

\bibitem{casson10}
Casson~F~J, Peeters~A~G, Angioni~C, Camenen~Y, Hornsby~W~A,
Snodin~A~P and Szepesi~G 2010
\newblock {\em Phys. Plasmas} {\bf 17} 102305.

\bibitem{barnes11}
Barnes~M, Parra~F~I, Highcock~E~G, Schekochihin~A~A, Cowley~S~C
and Roach~C~M 2011
\newblock submitted to \PRL

\bibitem{highcock10}
Highcock~E~G, Barnes~M, Schekochihin~A~A, Parra~F~I, Roach~C~M and
Cowley~S~C 2010
\newblock \PRL {\bf 105} 215003

\bibitem{navepc}
Nave~M~F~F 2010
\newblock personal communication

\bibitem{dorland00}
Dorland~W, Jenko~F, Kotschenreuther~M and Rogers~B~N 2000
\newblock \PRL {\bf  85} 5579

\bibitem{candy03a}
Candy~J and Waltz~R~E 2003
\newblock {\em J. Comput. Phys.} {\bf 186} 545

\bibitem{dannert05}
Dannert~T and Jenko~F 2005
\newblock {\em Phys. Plasmas} {\bf 12} 072309

\bibitem{peeters09c}
Peeters~A~G, Camenen~Y, Casson~F~J, Hornsby~W~A, Snodin~A~P,
Strintzi~D and Szepesi~G 2009
\newblock {\em Comput. Phys. Commun.} {\bf 180} 2650

\bibitem{parra10a}
Parra~F~I and Catto~P~J 2010
\newblock \PPCF {\bf 52} 045004
\\ Parra~F~I and Catto~P~J 2010
\newblock \PPCF {\bf 52} 059801

\bibitem{artun94}
Artun~M and Tang~W~M 1994
\newblock {\em Phys. Plasmas} {\bf 1} 2682

\bibitem{sugama97}
Sugama~H and Horton~W 1997
\newblock {\em Phys. Plasmas} {\bf 4} 405

\bibitem{parra08}
Parra~F~I and Catto~P~J 2008
\newblock {\em \PPCF} {\bf 50} 065014

\bibitem{hinton76}
Hinton~F~L and Hazeltine~R~D 1976
\newblock \RMP {\bf 48} 239

\bibitem{helander02bk}
Helander~P and Sigmar~D~J 2002 {\em Collisional Transport in
Magnetized Plasmas}
\newblock ({\em Cambridge Monographs on Plasma Physics})
\newblock ed Haines~M~G {\em et al}
\newblock (Cambridge, UK: Cambridge University Press)

\bibitem{parra11}
Parra~F~I, Barnes~M and Peeters~A~G 2011
\newblock {\em Phys. Plasmas} {\bf 18} 062501

\bibitem{simakov07}
Simakov~A~N and Catto~P~J 2007
\newblock \PPCF {\bf 49} 729

\end{thebibliography}
\end{document}